\documentclass[12pt,preprint]{aastex} 		

\newcommand{\kms}{km s$^{-1}$}

\newcommand{\etal}{et~al.}

\begin{document}

\title{Debris Disks of Members of the Blanco 1 Open Cluster\altaffilmark{1,2}}

\slugcomment{Accepted to ApJ} 
\shortauthors{Stauffer et al.} 
\shorttitle{Infrared Observations of Blanco 1}

\author{John R.\ Stauffer}
\affil{Spitzer Science Center, Caltech 314-6, Pasadena, CA  91125}

\author{Luisa M.\ Rebull}
\affil{Spitzer Science Center, Caltech , Pasadena, CA 91125}

\author{David James}
\affil{Physics and Astronomy Department, University of Hawaii at Hilo, Hilo, HI 96720}

\author{Alberto Noriega-Crespo}
\affil{Spitzer Science Center, Caltech , Pasadena, CA 91125}

\author{Steven Strom}
\affil{National Optical Astronomy Observatories, Tucson, AZ  85721 }

\author{Scott Wolk}
\affil{Harvard-Smithsonian Center for Astrophysics, 60 Garden St.,
Cambridge, MA  02138}

\author{Michael Meyer}
\affil{Steward Observatory, University of Arizona, Tucson,
AZ  85726}

\author{John M. Carpenter}
\affil{Astronomy Department, Caltech, Pasadena, CA 91125}

\author{David Barrado y Navascues}
\affil{Laboratorio de Astrofísica Espacial y Física Fundamental,
LAEFF-INTA, E-28080 Madrid, Spain}

\author{Giusi Micela}
\affil{INAF - Osservatorio Astronomico di Palermo Giuseppe S. Vaiana,
Piazza Parlamento 1, 90134 Palermo, Italy}

\author{Dana Backman}
\affil{Stratospheric Observatory for Infrared Astronomy and SETI Institute, 
    515 North Whisman Road, Mountain View, CA 94043}

\author{P. A. Cargile}
\affil{Department of Physics and Astronomy, Vanderbilt University, Nashville, TN 37235}

\altaffiltext{1}{This work is based (in part) on observations made
with the Spitzer Space Telescope, which is operated by the Jet 
Propulsion Laboratory, California Institute of Technology, under
NASA contract 1407. } 
\altaffiltext{2}{This publication makes use of data products 
from the Two Micron All Sky Survey, which is a joint project 
of the University of Massachusetts and the Infrared Processing 
and Analysis Center/California Institute of Technology, 
funded by the National Aeronautics and Space Administration 
and the National Science Foundation.}

\begin{abstract}

We have used the Spitzer Space Telescope to obtain Multiband Imaging
Photometer for Spitzer (MIPS) 24 $\mu$m photometry for 37 members of
the $\sim$ 100 Myr old open cluster Blanco 1.  For the brightest 25 of
these stars (where we have 3$\sigma$ uncertainties less than 15\%), we
find significant mid-IR excesses for eight stars, corresponding to a
debris disk detection frequency of about 32\%.   The stars with
excesses include two A stars, four F dwarfs and two G dwarfs.   The
most significant linkage between 24 $\mu$m excess and any other
stellar property for our Blanco 1 sample of stars is with binarity.  
Blanco 1  members that are photometric binaries show few or no
detected 24 $\mu$m excesses whereas a quarter of the apparently single
Blanco 1 members  do have excesses.  We have examined the MIPS data
for two other clusters of similar age to Blanco 1 -- NGC 2547 and the
Pleiades.  The  AFGK photometric binary star members of both of these
clusters also show a much lower frequency of 24 $\mu$m excesses
compared to stars that lie near the single-star main sequence.

We provide a new determination of the relation  between $V-K_{\rm s}$
color and $K_{\rm s}-[24]$ color for main sequence photospheres based
on Hyades members observed with MIPS.   As a result of our analysis of
the Hyades data, we identify three low mass Hyades members as
candidates for having debris disks near the MIPS detection limit.

\end{abstract}


\section{Introduction}
\label{sec:intro}

The past two decades have witnessed tremendous progress in
understanding circumstellar disk formation and early evolution. With
new results primarily from the Spitzer Space Telescope (Werner et al.\
2004) and its Multiband Imaging Photometer for Spitzer (MIPS; Rieke et
al. 2004) instrument,  we are beginning to understand the diversity of
old debris disk systems as a function of stellar mass (Rieke et al.\
2005; Beichman et al.\ 2006; Bryden et al.\ 2006; Carpenter et al.\
2009).  But the transition between  gas-rich primordial circumstellar
accretion disk and mature planetary systems such as our own remains
mysterious.  What fraction of young solar mass stars have debris
disks, and how does that fraction depend in detail on age?  How do the
properties of these disks depend on stellar metallicity or stellar
mass? Does debris disk formation or evolution depend on the birth
environment -- single versus binary; isolated versus rich cluster? Are
observations of debris disk systems consistent with expectations from
models of planet formation (e.g., Kenyon \& Bromley 2004)?

Determining why a given star has a detected disk while another
apparently similar star does not has proven difficult.   Probably the
clearest correlation is with age.  A number of  studies of relatively
young field stars or moving group members  (e.g., Rieke et al.\ 2005;
Su et al.\ 2006; Rebull et al.\ 2008; Carpenter et al.\ 2009) suggest
that the frequency for detection of debris disks around stars from a
few solar masses to about half a solar mass is largest at age
$\sim$10-30 Myr, and then slowly declines to older ages. Why stars of
similar age show diverse debris disk properties is much less clearly
constrained by observations.  It could be that the stars showing
excesses are primarily just the ones that have had recent, large
collisions in their asteroid belts producing copious amounts of small
dust particles via collisional cascades (Kenyon and Bromley 2004). 
Alternatively, stars with strong dynamo-driven winds may scour small
grains from their disks; the stars with detected IR-excesses might
therefore be those with comparatively weak winds (Chen et al.\ 2005;
Plavchan et al.\ 2005). Debris disks might have difficulty forming in
at least some types of binary systems, leading to a prediction that
the stars with mid-IR excesses might preferentially be single. 
However,  Trilling et al.\ (2007) obtained MIPS data for a sample of
A and F binaries, and concluded that on average, binary stars (in this
mass range) do not seem to have significantly different debris disk
detection frequencies compared to single stars.   For binary stars
that do have debris disks, the data suggested that debris disks are
most frequently seen for systems with close or ($<$ 3 AU) or wide
companions ($>$ 50 AU).

There are two complementary paths to address many of the above
issues: via observation of samples of nearby field stars and via
observation of stars in open clusters.   The field star path has the
considerable advantage that, in general, the targets can be selected
to be quite nearby, allowing smaller dust excesses to be detected and
allowing the photospheres of the star to be detected at 70 $\mu$m for
the nearest field stars. However, the field star sample cannot address
the question of dependence on stellar environment.  That is, while it
is known that most stars are born in clusters and subsequently become
field stars when their natal cluster is disrupted through dynamical
evolution (Lada et al.\ 1991; Porras et al.\ 2003), there is no way to
determine with any confidence whether a given field star was born in a
rich cluster or a small group.  Also, ages for field stars are very
uncertain and attempting to infer how debris disk characteristics
evolve with time using a field star sample is at best difficult.

The nearest open clusters offer the possibility to provide the best
answers to some of the disk evolution questions. Several programs have
obtained Spitzer MIPS data for members of the Pleiades (t $\sim$ 100
Myr; d $\sim$ 133 pc).  From a pointed survey as part of the Formation
and Evolution of Planetary Systems (FEPS; Meyer et al.\ 2006) Legacy
program, Stauffer et al.\ (2005) detected small 24 $\mu$m excesses
around $\sim$4 of 20 G dwarf members.  Independently of the FEPS
program, Gorlova et al.\ (2006) obtained a MIPS scan map of the
central square degree of the Pleiades, again detecting small 24 $\mu$m
excesses for about 15\% of the members.  Finally, Sierchio et al.\
(2010) obtained pointed 24 $\mu$m photometry for 37 solar type Pleiades
members, deriving a 32\% excess frequency for cluster members of about
solar mass. As a partial list, other open clusters for which MIPS data
have been obtained include:  IC 2391 (t $\sim$ 50 Myr; d $\sim$ 150
pc; Siegler et al.\ 2007);  NGC 2547 (t $\sim$ 30 Myr; d $\sim$ 390
pc; Gorlova et al.\ 2007);  NGC 2232 (t $\sim$ 25 Myr; d $\sim$ 350
pc; Currie et al.\ 2008); NGC2451 (t $\sim$ 50-80 Myr; d $\sim$
200-400 pc; Balog et al.\ 2009);  Praesepe (t $\sim$ 600 Myr; d
$\sim$ 180 pc; Gaspar et al.\ 2009); and NGC 2422 (t $\sim$ 100 Myr;
d $\sim$ 425 pc; Gorlova et al.\ 2004).  

\begin{figure} %
     \epsscale{0.8}
     \plotone{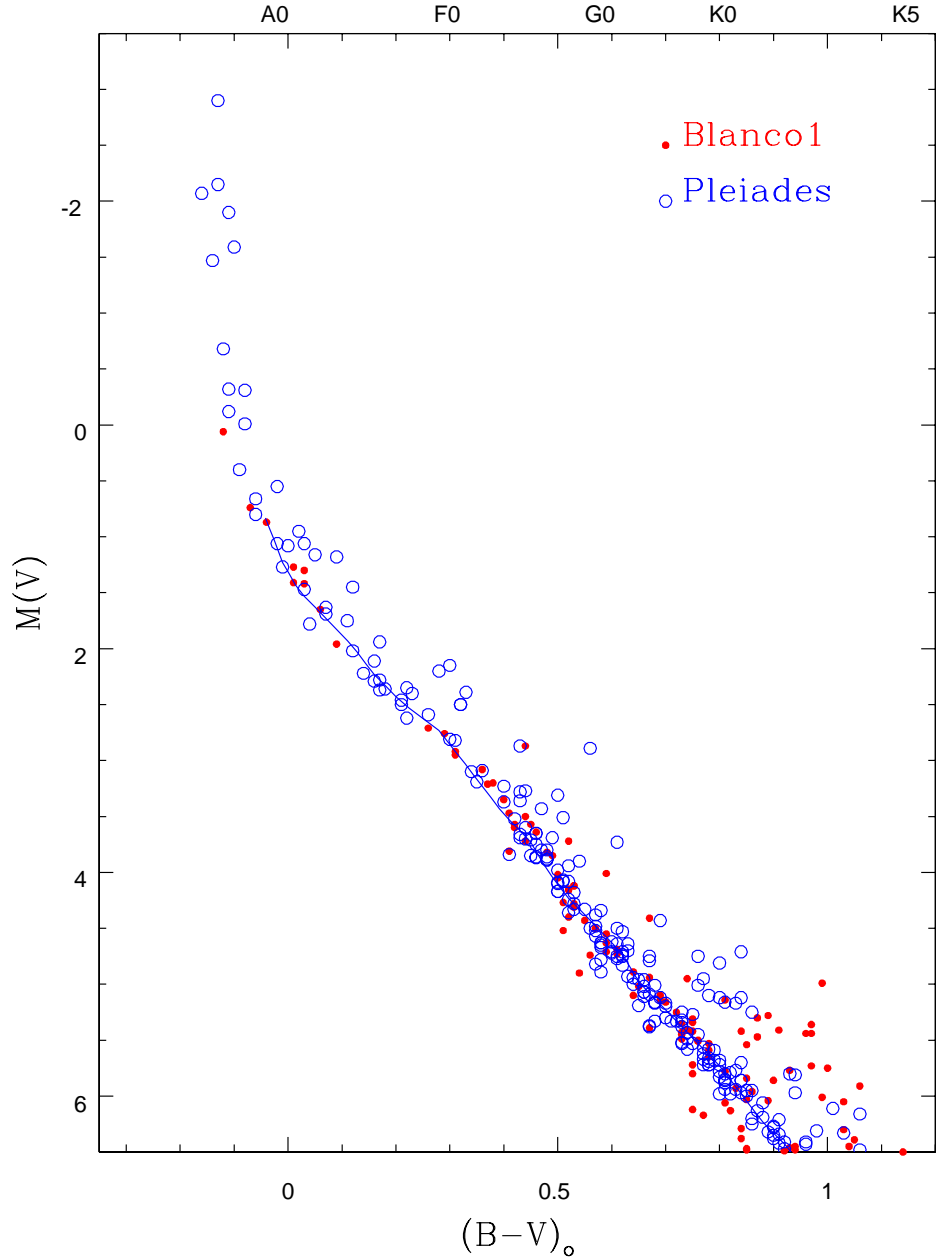}
    \caption{Color-magnitude diagram for high mass members of the
Pleiades and Blanco 1.  Distances of
133 and 250 pc, and $E(B-V)$ of 0.04 and 0.01 for the Pleiades and
Blanco 1, respectively, were assumed.   Blanco 1 is deficient in
high mass stars compared to the Pleiades.
    \label{fig:blanco_imf}
    }
\end{figure}

\begin{figure} %
     \epsscale{0.8}
     \plotone{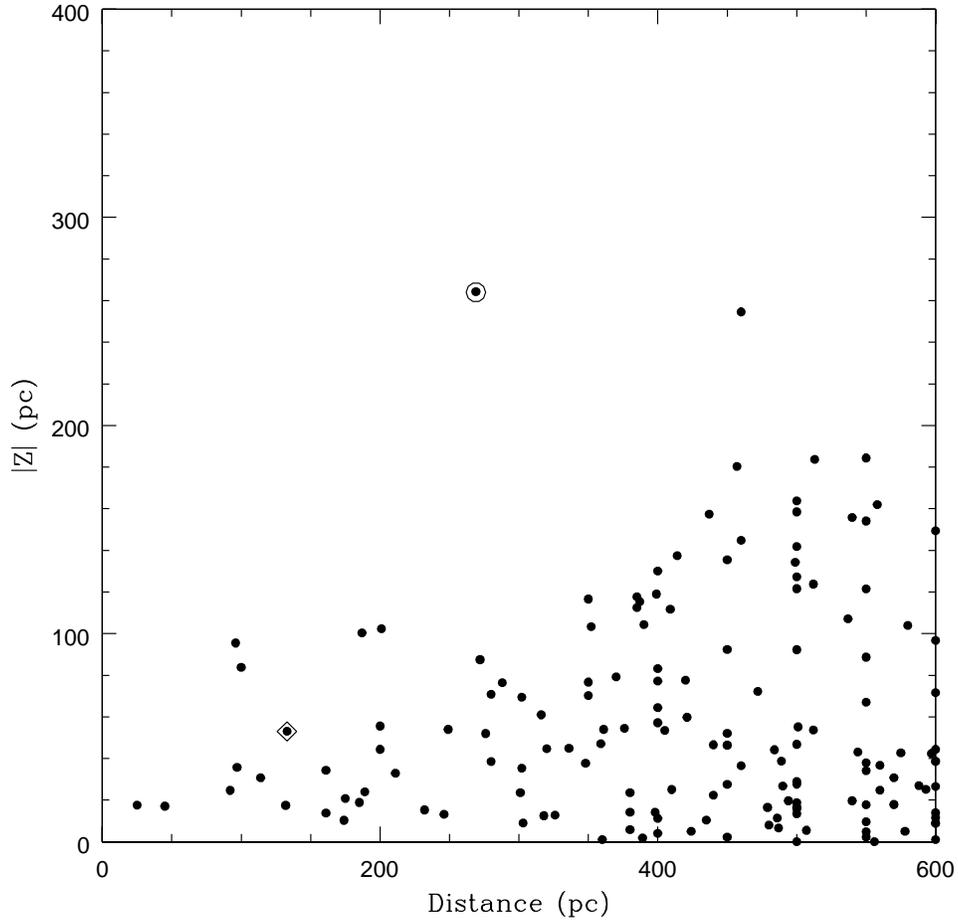}
    \caption{Height above the galactic mid-plane versus distance for
all open clusters in the Dias et al.\ (2002) catalog.  The open circle
marks Blanco 1; the diamond symbol marks the Pleiades.   Blanco 1
is located further above the mid-plane than any other cluster within
this distance range.
    \label{fig:blanco_height}
    }
\end{figure}

Open clusters come with a wide range of initial properties -- at least
richness and metallicity, but potentially also the IMF, binary
fraction or primordial disk fraction might vary from cluster to
cluster. Now that open clusters have been observed that sample at
least most of the age range that is of interest to debris disk
evolution studies, it would seem of interest to observe clusters that
could possibly shed light on whether debris disk frequency depends in
any significant way on cluster properties other than age.  As noted
above, the Pleiades provides a template for debris disk properties at
100 Myr for a rich, solar-metallicity, kinematically normal cluster
(i.e. the space motion of the Pleiades is very similar to the space
motion of the many nearby, young stars that form the Local Association).  
Another relatively nearby open cluster with an age of $\sim$100 Myr
is Blanco 1 (galactic latitude $\sim -79$ degrees;
ecliptic latitude = $-28$ degrees; distance = 250 pc; $E(B-V)$ =
0.01). As illustrated in Figures~\ref{fig:blanco_imf} and
\ref{fig:blanco_height},  Blanco 1 has some properties that make it
quite dissimilar to the Pleiades. First (Fig.~\ref{fig:blanco_imf}),
Blanco 1 is much less rich, but is particularly deficient in high mass
stars -- specifically, it has just two B star members (a B8 and a B9)
and no member with M(V) $<$ 0.0 versus the Pleiades having 14 B stars
(and at least one white dwarf member, whose precursor was presumably
an O or early B star).   At very early ages ($<$ a few Myr) therefore,
dynamical interactions (Bate et al.\ 2003) 
or the winds and UV radiation from high mass stars (Adams et al.\ 2004)
may have truncated the primordial disks of the low mass stars in
the proto-Pleiades more than in the proto-Blanco 1 cluster. 
Figure~\ref{fig:blanco_height} shows that Blanco 1 is an extreme
outlier in terms of its height above the galactic plane.  Either the
molecular cloud from which it formed had unusual kinematics (it was a
high-velocity cloud) or some external event, such as a very nearby
supernova or winds from nearby O stars, imparted a significant amount 
of momentum to the proto-cluster when it was being formed (Oort \&
Spitzer 1955).  Radiation driven implosion models for star-formation
triggered by external events in some cases predict significantly
high accretion rates for protostars (Motoyama et al.\ 2007) or IMFs
different from normal star-formation events (Bertoldi 1989).

If everything involved in debris disk formation and evolution only
involves the local environment of individual protostars, then one
would predict no significant difference in the debris disk frequency
and IR-excess sizes of Pleiades and Blanco 1 members. If the debris
disk frequency in Blanco 1 were significantly different from that of
the Pleiades, explanations involving the differing properties of the
two clusters could be explored which might lead to a better general
understanding of what controls debris disk formation and evolution.  
In this paper, we present MIPS 24 $\mu$m data for a sample of about 40
members of  Blanco 1.  Our goals are to better understand the  (a)
disk frequency at 100 Myr; (b) how the debris disk properties of the
Blanco 1 members compare to those for other similar age open clusters;
and (c) to examine how debris disk presence correlates with other
stellar  properties such as rotation, X-ray luminosity  or binarity.

\section{Target Selection and Spitzer Observations}
\label{sec:selection}

Based on very recent efforts, good membership lists for Blanco 1 now
exist over essentially the entire stellar mass range (Moraux et al.\
2007; Mermilliod et al.\ 2008; Cargile et al.\ 2009; Gonzalez \&
Levato 2009; James et al.\ 2010).  However, when the target list for
this program was constructed in 2004, membership information was much
spottier.  The primary information available to us included several
large photometric studies (Westerlund et al.\ 1988; Edvardsson et al.\
1995),  moderate resolution spectroscopy providing chromospheric
diagnostics and lithium equivalent widths (Panagi et al.\ 1994, 1997),
and X-ray imaging of the cluster to identify stars with strong coronal
emission (Pillitteri et al.\ 2004 and references therein).

The twenty-five stars we selected for observation with MIPS consist
of a nearly complete list of the high probability members of the
cluster known in $\sim$2004, within the approximate spectral type
range A0 to G0.  All of these stars were selected as photometric and
proper motion members of the cluster; most of the later-type stars
also have activity indicators supportive of membership.

With new membership available to us now based on modern  $BVIJHK_{\rm
s}$ photometry, accurate radial velocities and improved proper motions
(Mermilliod et al.\ 2008, Gonzalez \& Levato 2009), we have
re-examined the membership of these stars.  All of them are confirmed
as high-probability cluster members in one or both of Mermilliod et
al. or Gonzalez \& Levato, with all of them being radial velocity
members of the cluster based on multiple observations.  All of them
are also confirmed as cluster members from proper motion studies, and
all of them are good photometric members of the cluster (see Figures 1
and 2).

The targeted cluster members, their J2000 positions, and the Spitzer
Astronomical Observation Request (AOR) identification numbers
(AORKEYs) for our MIPS observations are provided in Table 1.  We
generally provide two identification names for our targets -- the name
from Westerlund et al.\ (1988) [Wxxx] and the name from Epstein \& de
Epstein (1985) [ZSxxx, where `ZS' stands for Zeta-Sculptor, the
alternative name for the Blanco 1 cluster, derived from the brightest
proposed cluster member - although Zeta Zcl is no longer considered a
member of the cluster]. These data were obtained under Spitzer program
30022.

Each target was observed using a MIPS photometry AOR at 24 and 70
$\mu$m.  Our AORs are the same for all of our targets and are very
similar to those used for most of the FEPS Pleiades observations. In
particular, we obtain four cycles of 10 seconds integration time for
24 $\mu$m and 7 cycles of 10 second integration time for 70 $\mu$m. We
limited ourselves to 7 cycles at 70 $\mu$m based on Fig.\ 7 of Bryden
et al.\ (2006), which showed that the confusion noise level was
reached for typical galactic fields after about five cycles of 10
second integration time photometry for MIPS-70.  None of the targets 
were  detected at 70 $\mu$m, and the upper limits were deemed to be
too shallow to be of use in the current study, and so we do not
discuss 70 $\mu$m data any more in the present paper. 

Each of our MIPS observations produced two regions of the sky covering
about $6^{\prime}\times6^{\prime}$ with flux densities at 24 $\mu$m --
one centered on the target object, and a second serendipitous field
offset from that primary field by of order 12$^{\prime}$.  In order to
attempt to expand the number of stars in our study, we also
cross-correlated a master list of possible Blanco 1 cluster members
against the area of the sky included within our MIPS coverage. This
master list is a merger of the candidate member catalog from Panagi et
al.\ (1997) and Table 1 from Mermilliod et al.\ (2008).  We also
incorporated membership information for the brightest cluster members
from Perry et al.\ (1978) and from the Hipparcos proper motions
(Perryman et al.\ 1997).  Additional membership data for faint
candidate cluster members came from new photometry, proper motions,
radial velocities, and lithium equivalent widths reported in James et
al.\ (2010).  Twenty additional probable Blanco 1 members were
identified which are detected in our MIPS images.  However, we were
only able to derive reliable flux densities for twelve of these stars (the
ones brighter than $K_{\rm s}$ = 11). These twelve stars are almost
all fainter (and hence lower mass) than the primary targets.  Nine of
these additional members are high probability cluster members based on
the Mermilliod et al.\ (2008) program (i.e., based on proper motions,
photometry and multiple radial velocity measurements).  Of the
remaining sources, ZS38 is a proper motion and photometric member, but
is also an X-ray source (Cargile et al.\ 2009) and a spectroscopic
member based on radial velocity and lithium abundance (Jeffries \&
James 1999).  ZS83 is a photometric member, a proper motion member,
and an X-ray source (Cargile et al.\ 2009); ZS83 has a lithium
abundance N(Li) = 2.3 which for its $\sim$K0 spectral type indicates
it is quite young (Jeffries \& James 1999).  The Jeffries \& James
radial velocity for ZS83 is a few kilometers per second off the
cluster mean, which Cargile et al.\ (2009) take to indicate it is an
SB1.  Given the combination of proper motion, X-ray detection, and
youth based on lithium, we agree and conclude ZS83 is a highly
probable member of the cluster.  That only leaves ZS108 as having no
published, modern data supportive of cluster membership.  In James et
al.\ (2010) we show that it too is a probable cluster member, based on
its photometry ($V$ = 13.46; $B-V$ = 1.06; $V-I_c$ = 1.21), proper
motion ($\mu_{RA}$ = 23.4 $\pm$ 0.5 mas yr$^{-1}$; $\mu_{DEC}$ = 3.9
$\pm$ 0.8 mas yr$^{-1}$; membership probability P = 57\%), and lithium
abundance (EqW = 22 m\AA; N(Li) $\sim$ 0.6).   James et al.\ obtained
six WIYN/HYDRA spectra of ZS108, one of which showed it to be an SB2
and the others show it to be a radial velocity variable (ranging from
$-$20 to +40 \kms).  Based on these data, we consider it to be a
probable member of the cluster.  A complete list of the Blanco 1
members for which we have obtained MIPS photometry is provided in Table 2.

\section{MIPS Photometry}
\label{sec:observations}

We extracted the individual Basic Calibrated Data (BCD) files
(individual flux-calibrated array images) from the Spitzer archive,
and then used the Spitzer Science Center (SSC) MOPEX package (Makovoz
\& Marleau 2005) to construct mosaics for each AOR.   We used the
default interpolation scheme and outlier rejection for MOPEX, which is
discussed in Makovoz and Marleau (2005) and the online MOPEX help.  We
self-flattened following the SSC recommendations, using IRAF. We
overlap-corrected using the default settings in the overlap module
distributed as part of MOPEX.  Using the APEX-1Frame package from
MOPEX, we performed PRF fitting to obtain photometry for each object
at 24 $\mu$m using an empirical PRF derived from photometry mode
observations of relatively bright stars in relatively high galactic
latitude fields (i.e., from data as similar as possible to the Blanco
1 observations). We also obtained aperture photometry for all the
cluster members, using the PHOT package in IRAF.  We employed an
aperture of 2 pixel radius, with the sky annulus from 7 to 11 pixels. 
We used an aperture correction factor of 0.716 magnitudes derived
from bright stars.  We
converted each image to DN prior to doing the aperture photometry
(using the DN sec$^{-1}$ MJy$^{-1}$ sterradian$^{-1}$ value in the
FITS headers) in order to allow use of the photometric uncertainties
calculated by the PHOT package.   Brighter than [24] = 10, the two
flux density determinations generally agree to within a couple
percent, and the values we report are the average (except as noted
below);  for fainter stars, the aperture flux densities are noisier
and we report only the PRF flux densities for these stars. Based on
the calibration information provided by the SSC, the flux density for
zero magnitude for the MIPS 24 micron magnitude scale is 7.14 Jy.

For the stars that were our original targets, each star is centered in
the 24 micron array FOV, with each star falling on or near the same
pixels and being observed with the standard MIPS photometry mode AOR. 
This is the same observing pattern as used for the MIPS calibration
stars.  Therefore, compared to stars that fall in other parts of the
MIPS FOV (or compared to stars observed in MIPS scan map mode) some
possible sources of error in the photometry are eliminated/avoided.  
For MIPS calibration stars with essentially arbitrarily large
intrinsic S/N, repeat observations show an RMS scatter in the measured
24 $\mu$m flux densities of order 0.5 to 0.7\%, showing that the MIPS 24
$\mu$m camera is stable enough to deliver  1\% photometry if the
target star is bright enough (Engelbracht et al.\ 2007). The
sensitivity prediction tool for MIPS predicts SNR $>$ 100 for our
brightest targets, and SNR of a few percent for our faintest, targeted
stars.  The uncertainty estimates returned by PHOT agree with these
expectations, and we report these uncertainties (for our original
target stars) as the parenthetical numbers following the 24 $\mu$m
magnitudes in Table 2.

For the serendipitous Blanco 1 members reported in Table 2, other
sources of error may affect the photometry (flat field errors; array
location dependent PSF or filter response corrections). Rather than
adopt the uncertainties for each star reported by APEX or PHOT, we
have chosen to determine empirical uncertainty estimates for stars of
a given flux density using regions of the sky where we have
overlapping MIPS coverage (i.e., where the mosaics from two or more
AORs overlap). We have two or more observations for more than 100
point sources which we sorted by flux density and grouped into several
flux density bins.  Within each group, we calculated the mean absolute
difference for the two measures; we converted the mean absolute
difference to an estimate of the error ($\sigma$) using the standard
formula ($\sigma = 1.2\times$MAD).  The derived 1-$\sigma$ values
range from 4\% for our brightest members to  about 10\% at K$_{\rm s}$
= 11.   We report these uncertainties in Table 2 for the serendipitous
cluster members (the last twelve stars in the table).

Two of the original target stars (W58 and W60) have ``companions"
within 2 or 3 pixels in the MIPS-24 images of comparable (though
fainter) magnitude.  Neither companion is visible in 2MASS images,
making it likely these are IR-bright galaxies.  The aperture
photometry for these stars is compromised by the proximity of the
companion object.  We therefore report the PRF-fitting magnitude for
these sources.

\begin{figure} %
     \epsscale{0.8}
     \plotone{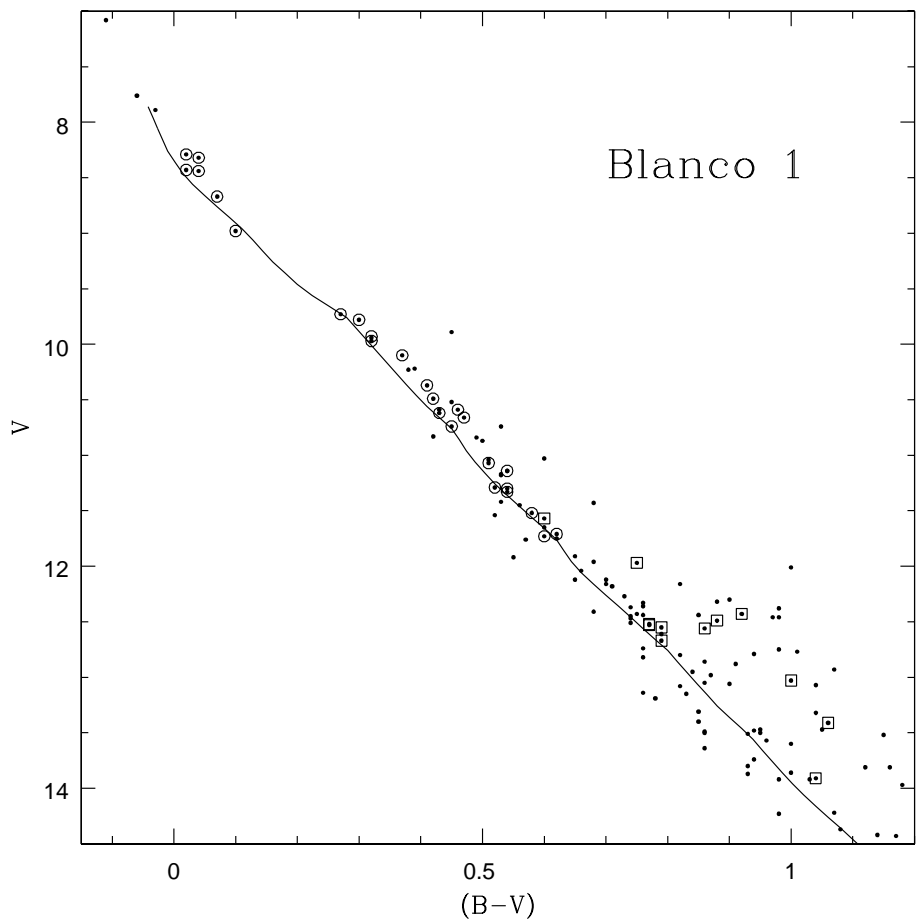}
    \caption{Color-magnitude diagram for stars in our Blanco 1 master
catalog and the stars for which we have obtained MIPS photometry.  The
curved line is a fit to the Pleiades single-star main sequence locus,
as provided in Stauffer et al.\ (2007).  We have assumed distances of
133 and 250 pc, and $E(B-V)$ of 0.04 and 0.01 for the Pleiades and
Blanco 1, respectively.   The stars we targeted in Blanco 1 are shown
with circles; the additional members that happen to fall within the
area covered by our MIPS mosaics are shown as open squares.
    \label{fig:blanco1bmv_cmd}
    }
\end{figure}

\begin{figure} %
     \epsscale{0.8}
     \plotone{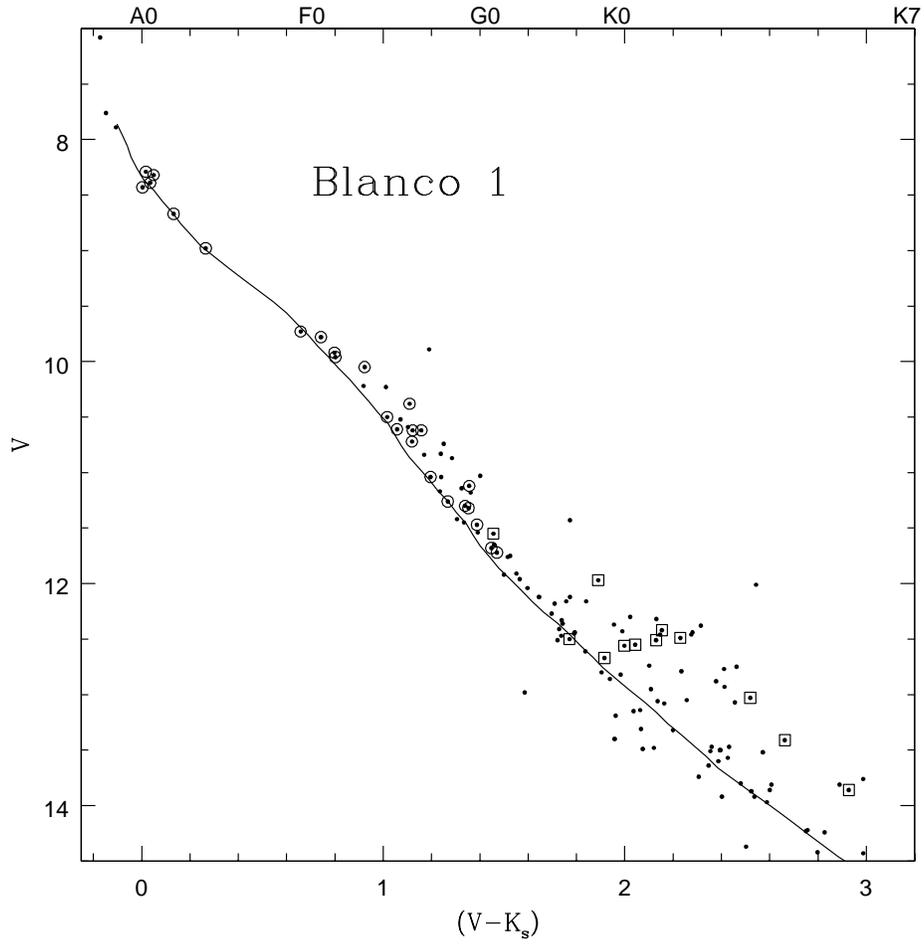}
    \caption{As for Figure 3, except using $V-K_{\rm s}$ as the abscissa.
    \label{fig:blanco1vmk_cmd}
    }
\end{figure}

Table 2 provides data for the entire set of cluster members for which
we have obtained MIPS-24 flux densities.  
Figures~\ref{fig:blanco1bmv_cmd} and \ref{fig:blanco1vmk_cmd}  show
color-magnitude diagrams using $B-V$ and $V-K_{\rm s}$ as the abscissa
for our entire catalog of candidate cluster members, with the stars
observed with Spitzer highlighted (circles for our original targets;
squares for the fainter, serendipitous members).   The primary
reference for membership in the cluster for each star is indicated by
first reference given in the right-most column of Table 2.  If only
one reference is given, that reference is also the source for the
optical photometry.  If other references are given, they provide the
$V$ and $B-V$ (``NS" refers to NStED -- the NASA Stellar and Exoplanet
Database; for the Blanco 1 stars, the NStED $V$ and $B-V$ photometry
come from Tycho-2, where the native Tycho photometry have been
converted to Johnson $B$ and $V$ using a formula from Bessell 2000). 
The K$_{\rm s}$ photometry is from the 2MASS point source 
catalog (Skrutskie et al.\ 2006).
The spectroscopic rotational velocities are nearly all from Mermilliod
et al.\ (2008); exceptions are those for ZS38 and ZS102 -- from
Jeffries \& James\ (1999) -- and those for W35, W96, W104 and ZS83 --
from James et al.\ (2010). The rightmost column of the table also
identifies stars whose binarity have been empirically detected via
radial velocity variability (SB or SB2) or as a close visual pair (VB)
-- these identifications come from Mermilliod et al.\ (2008) except
for ZS38 and ZS108 which are attributed to James et al.\ (2010).  

The X-ray luminosities in Table 2 come from Micela et al.\ (1999),
Pilliteri et al.\ (2003; 2004), and  Cargile et al.\ (2009).  Because
it is the most recent reference, we give preference to the values
provided by Cargile et al.  The letters following L$_x$ in the table
point to the source of the X-ray data.   Where only X-ray luminosity
was provided, we have calculated log(L$_x$/L$_{bol}$) ourselves, using
bolometric luminosities provided by Schmidt-Kaler (1982). The
literature values for L$_x$ were calculated using slightly different
assumptions (distance of 240 or 250 pc; $E(B-V)$ =  0.016 or 0.02) --
we have ignored those small differences. 

Our sample of Blanco 1 stars with MIPS  photometry 
extends over the spectral type range from A0 to about mid-K.

\section{Use of Spitzer Photometry of the Hyades to Calibrate the
Photospheric $K_{\rm s} - [24]$ color of Low Mass Stars} 
\label{sec:calibration}

Following the path forged by several other groups (e.g., Rebull \etal\
2008 and references therein), as the measure of IR excess for the
Blanco 1 cluster stars, we will use $K_{\rm s} - [24]$ color, rather
than fitting a photospheric model to the shorter wavelength photometry
and comparing the measured 24 $\mu$m flux density to the predicted
photospheric flux density.   We believe that this type of analysis can, in
fact, provide more sensitive detection limits for debris disks than
SED analysis because it introduces the fewest possible assumptions and
model dependencies.  Also, the fact that the 2MASS catalog provides
homogenous, sensitive $K_{\rm s}$-band photometry over the entire sky
insures that our measured IR-excess parameter can be placed on a
common system with other stellar samples to high accuracy.  However,
in order to obtain the most benefit from using this method, it is
necessary to have a good calibration of the photospheric colors in
this plane.  

For early type stars, $K_{\rm s} - [24]$ color is essentially zero,
making it easy to identify stars with excesses.   However, as
demonstrated by Gautier et al.\ (2007), the photospheric $K_{\rm s} -
[24]$ departs from zero as one goes to later spectral types, reaching
values of several tenths of a magnitude at early M. Gorlova et al.\
(2007) derived a calibration of the trend of $K_{\rm s} - [24]$ color
with $V-K_{\rm s}$ based on members of the Pleiades which is valid for
the mass range of our Blanco 1 sample. However, the Pleiades is
relatively far away and nebular emission adds structured noise,
resulting perhaps in a less well-defined relation than desirable.   We
decided to investigate whether the Spitzer archive provides a better,
empirical means to define the photospheric $K_{\rm s} - [24]$  color
for low mass stars. 

\begin{figure} %
     \epsscale{0.8}
     \plotone{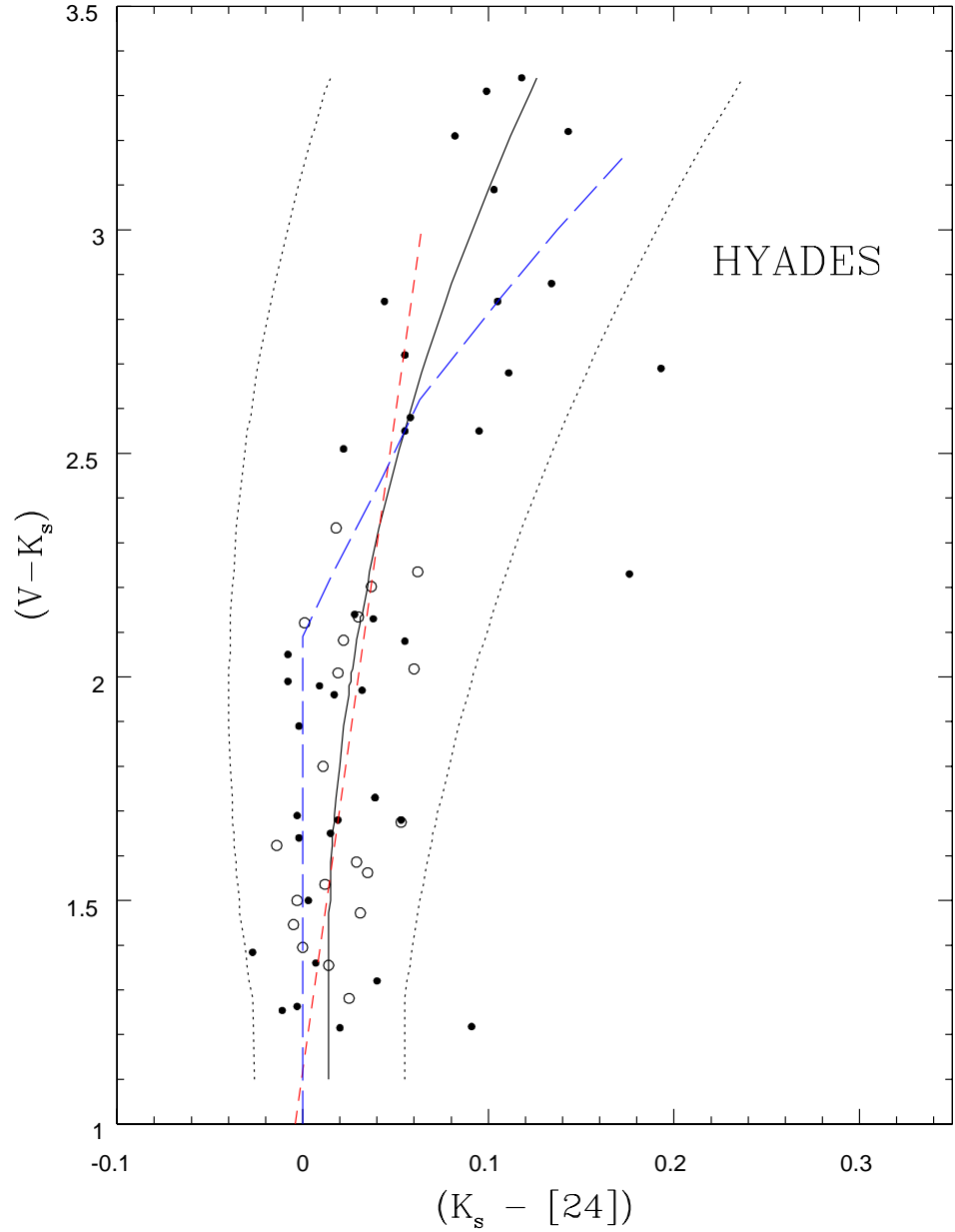}
    \caption{MIPS-24 photometry for members of the Hyades, illustrating
how the photospheric $K_{\rm s} - [24]$ color becomes redder for late type dwarfs.
The solid curve is a fit to these data, excluding the three apparent outliers.
Open circles are from PID148 and filled circles are from PID3771.  The red,
dashed line is the relation from Gorlova et al.\ (2007); the blue, dashed
line is Plavchan et al.\ (2009) relation - see text for details. 
    \label{fig:hyad_km24}
    }
\end{figure}

After some experimentation, we believe that Spitzer observations for
the Hyades provides the best data set to define the photospheric
$K_{\rm s} - [24]$  relation.  The Hyades  has been well-observed by
Spitzer, it is the closest open cluster (and hence has the best
signal-to-noise), and data exist for essentially the entire spectral
type range for which we have Blanco 1  observations.   The Spitzer
observations of the Hyades were obtained as part of two programs --
PID 148 (Carpenter et al.\ 2008)  and PID 3371 (Cieza et al.\ 2008). 
Except where mentioned, we have simply used the Spitzer flux densities
published in those papers. Both papers used the same flux density for
zero magnitude for MIPS 24 $\mu$m as we have, and both adopted K$_{\rm
s}$ magnitudes from 2MASS. Figure~\ref{fig:hyad_km24} shows how
$K_{\rm s} - [24]$ color varies with $V-K_{\rm s}$ color over the
spectral type range from G0 to M0.  The solid curve is our fit to
these data, corresponding to: \begin{equation} K_{\rm s} - [24] =
+0.042 - 0.053\times(V-K_{\rm s}) + 0.023\times(V-K_{\rm
s})^2\end{equation}   The dashed curves are the 3-$\sigma$ bounds
around the fiducial curve, where we have calculated a running estimate
of the RMS by ordering the stars by their $V-K_{\rm s}$ color, and
determining the RMS for each group of 10 successive stars.  The
3-$\sigma$ bound ranges from about 0.06 mag at $V-K_{\rm s}$ = 1.5 to
about 0.10 mag at $V-K_{\rm s}$ = 3.0. We exclude one observed star
from this plot -- HD 242780 -- because we discovered it had been
observed two hours after a MIPS 24 $\mu$m observation of Mon R2 which
badly saturated the array and left it in an unstable state at the time
of the HD 242780 observation. Another star, HD 30505, appeared to have
a 24 $\mu$m excess using the published photometry from Cieza et al.\
(2008).  However, our own flux density measurement is about 10\%
fainter than the published value, and quite consistent with other
Hyades stars of the same spectral type; we plot our flux density for
this star.  Our flux densities for a sample of a half dozen other
stars from Cieza et al.\ agree with their values within the expected
uncertainties; also, the Cieza et al.\ and Carpenter et al.\ stars
show no significant systematic differences in their location in
Figure~\ref{fig:hyad_km24}.

Also shown for reference in Figure~\ref{fig:hyad_km24} are the 
Gorlova et al.\ (2007) and Plavchan et al.\ (2009) predictions for the
photospheric $K_{\rm s} - [24]$ color, where for the Plavchan et al.\
relation we have converted $T_{\rm eff}$ to $V - K$ using the
color-temperature  data in Kenyon \& Hartmann (1995).  Our relation
differs only slightly from the Gorlova et al. predictions over the
range of validity of the two relations.  We differ more from the
Plavchan et al. relation, particularly at the red end.  The bluest
field M dwarfs of Gautier et al.\ (2007) have $K_{\rm s} - [24]$
$\simeq$ 0.35 at $V-K_{\rm s}$ = 4.  The Plavchan et al. relation is
designed to link smoothly to the field M dwarfs; our Hyades relation
would have to steepen significantly just redward of our faintest stars
to be consistent with the Gautier et al. data.  For the purposes of
this current paper where all of the objects of interest have $V-K_{\rm
s}$ $<$ 3, we believe the Hyades relationship provides the best
available means to predict the photospheric $K_{\rm s} - [24]$ color.

There are three Hyades members that fall outside the 3-$\sigma$ upper
limit shown in Figure~\ref{fig:hyad_km24} -- they are vB 19 (HD
26784), VA 133 (HDE 285690),  and VA 407 (HDE 286789) -- all from PID
3771.  We have made our own measurement of the 24 $\mu$m flux density
for these three stars.  Our flux densities are about 3\% fainter than
the published ones for VA 133 and VA 407, but even using our values,
these objects' $K_{\rm s} - [24]$  color falls above the 3-$\sigma$
curve.  Our flux density for vB 19 is essentially the same as the
published value.  We see no evidence in the MIPS nor 2MASS images for
any source confusion.  Cieza et al.\ (2008) did not identify these
stars (or any other low mass Hyades member) as debris disk
candidates.  Their method to identify excesses was to use NextGen
(Hauschildt et al. 1999) model atmosphere predictions of the IR SEDs
-- using published spectral types of the Hyades stars to select which
model to use, and normalizing the model spectra to their 2MASS
photometry.  Because our color-based technique for estimating excesses
is internally self-calibrating (the majority of the stars define the
locus of non-excess objects) and has fewer additional input parameters
(each of which may introduce further uncertainties), we believe our
technique is capable of detecting smaller excesses, and that these
three stars have small IR excesses.

\section{Identification of Blanco 1 Members with Apparent 24$\mu$m Excesses
and the Physical Interpretation of Those Excesses}
\label{sec:analysis}

\subsection{Blanco 1 Members with Excesses}

\begin{figure} %
     \epsscale{0.8}
     \plotone{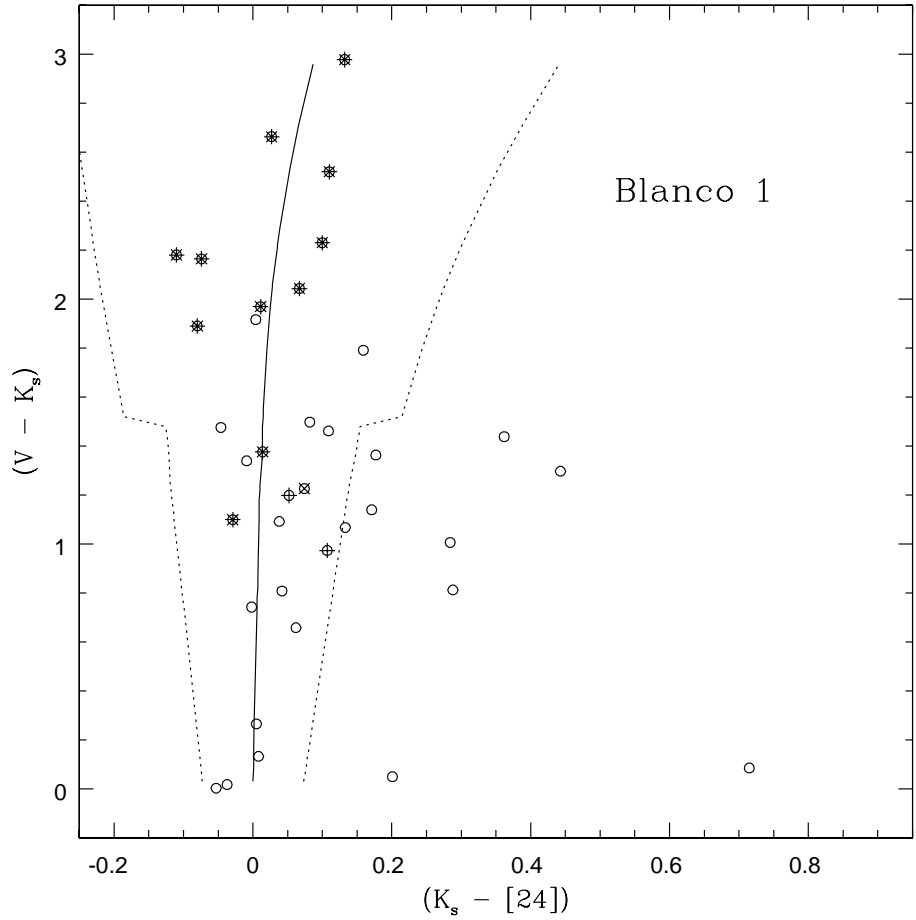}
    \caption{MIPS-24 photometry for members of Blanco 1. 
    The solid curve is the expected photospheric relation as derived from
  the Hyades low mass stars, as shown in the previous figure.  The dashed
  curves are the three sigma limits around this curve, as discussed in the
  text.  Objects identified as spectroscopic or visual binaries are overplotted
  with a cross; objects identified as photometric binaries are overplotted
  with a plus sign; objects identified by both techniques then appear as
  the combination of symbols.
    \label{fig:blanco_km24}
    }
\end{figure}

Figure~\ref{fig:blanco_km24} shows a plot of $K_{\rm s} - [24]$ color
versus $K_{\rm s}$ magnitude for the members of Blanco 1 for which we
have MIPS observations.  The solid curve is the relation derived from
the Hyades stars (for $V-K_{\rm s} >$ 1.2); for $V-K_{\rm s} <$ 1.2,
we simply extrapolate this relation to $V-K_{\rm s}$ = 0.0 where we
adopt $K_{\rm s} - [24]$ = 0.00 at that color. The Blanco 1
3-$\sigma$\ curves are based on our estimated 24 $\mu$m\ flux density
uncertainties as described in \S3.  Specifically, for $K_{\rm s}$ $<$
10, we adopt one sigma $K_{\rm s} - [24]$ uncertainties ranging from
0.02 at $K_{\rm s}$ = 8 to 0.045 at $K_{\rm s}$ = 10; for the fainter
stars,  the adopted one sigma $K_{\rm s} - [24]$ uncertainties range
from 0.07 at $K_{\rm s}$ = 10 to 0.10 at $K_{\rm s}$ = 11.   

Eight stars fall well to the right of the 3$\sigma$  upper limit curve in
Figure~\ref{fig:blanco_km24}. They are W23 (A0V), W28 (F5), W35
(A9/F0), W38 ($\sim$F5), W53 ($\sim$G0),  W88 (A0V), W91 ($\sim$G0) and 
W99 ($\sim$F8).  W60 ($V - K_{\rm s}\sim$ 1.06, $K_{\rm s}$ = 0.13)
falls just slightly above the 3$\sigma$ limit, however, we do not
include it as a detected excess star because it is one of the two
cluster members with nearby sources contaminating the MIPS photometry
(see previous section), thus making its photometry more uncertain.
Assuming all eight of the stars with 24 $\mu$m excesses have debris
disks, the derived disk frequency for our sample of A0 through M0
Blanco 1 stars is 8/37 or about 22$\pm$8\%. However, because our [24]
uncertainties are relatively large for the fainter, serendipitous
members, if we restrict ourselves to just the original set of targeted
cluster members where our uncertainties allow us to detect 3$\sigma$
excesses of 0.15 mag or larger, our detection frequency is 8 of 25, or
32$\pm$11\%.  This is comparable to the 30-45\% detection frequency for
NGC2547 B-F stars found by Gorlova et al.\ (2007) and the 32$\pm$7\%
detection frequency for late F to early K dwarf members of
the Pleiades by Sierchio  et al.\ (2010).  If one restricts the
comparison to just the Pleiades and NGC2547 stars in about the same
mass range as the Blanco 1 stars (see \S 6 and the appendix), the Pleiades
and NGC2547 disk fractions are 21$\pm$4\% and 41$\pm$7\%, respectively.

\subsection{AGN Contamination?}

Both active galaxies (Seyferts and QSOs) and star-forming galaxies are
bright at 24 $\mu$m.  The surface density of extragalactic sources
increases rapidly at fainter flux densities (Papovich et al.\ 2004). 
As one attempts to detect debris disks around progressively lower mass
members of open clusters, the probability that the line-of-sight to a
given cluster member intersects a comparably bright distant galaxy
increases. Fainter than [24] = 11 (correponding to 24 $\mu$m flux
density $\sim$ 0.3 mJy), the chance that a 15\% flux density excess is
due to a random line-of-sight contamination by an extragalactic source
becomes quite significant.   That was one reason we chose to not
report 24 $\mu$m photometry for the faintest detected sources in
Blanco 1.  One can use the source counts in Papovich et al.\ (2004) to
estimate the probability that any of our reported debris disk
candidates are false positives due to extragalactic contamination.  
We adopt the same assumptions for our contamination estimates as in
Gaspar et al.\ (2009) for their analysis of Spitzer MIPS observations
of Praesepe.  Specifically, we adopt a matching radius for chance
alignment of 3.6 arcseconds and a minimum flux density excess of 15\%
of the photospheric flux density.   For our seven brightest cluster
members (7.5 $<$ [24] $<$ 9), the surface density of sufficiently
bright extragalactic sources is relatively low, and the probability
that any of these sources has a 24 $\mu$m excess to line-of-sight
contamination is small ($<$ 5\%).  However, given the total number of
targets in our sample with 9 $<$ [24] $<$ 11, one would expect more
than a 50\% chance of finding one falsely identified 24 $\mu$m excess
source (versus the five objects in this mass range that we actually
identify as having excesses). In the event of such a contaminant, in
most cases the galaxy would not be $\it{exactly}$ aligned with the
Blanco 1 member, and the resultant 24 $\mu$m centroid would be
displaced from the true Blanco 1 stellar position (as defined by the
position derived from the 2MASS data).  Our debris disk candidates
show no such positional offsets, and we therefore prefer to accept all
six as having real 24 $\mu$m excesses.

\subsection{Correlation of IR-excess and Other Stellar Parameters}

\begin{figure} %
     \epsscale{0.8}
     \plotone{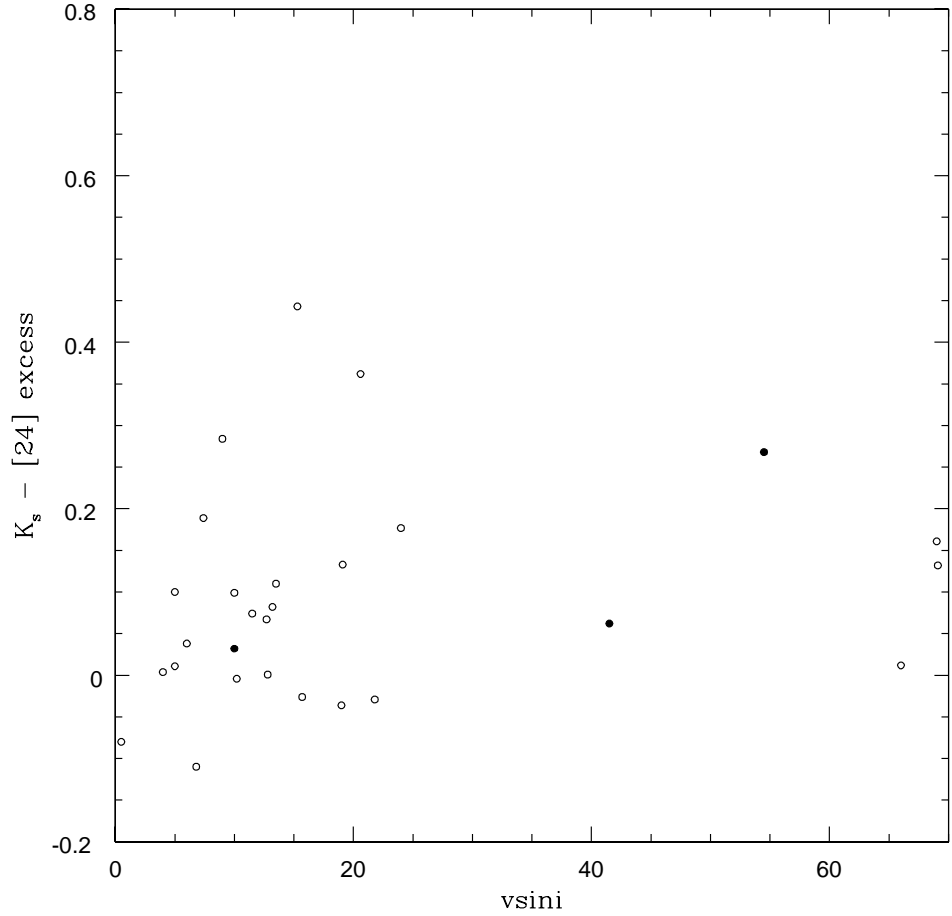}
    \caption{$K{_s}$ - [24] color versus spectroscopic rotational velocity
for members of Blanco 1 for which we have MIPS data.  Filled dots are early
type stars with $B-V$ $<$ 0.35 (corresponding to spectral type earlier than
about F3), which have no outer convective envelope and hence for which there
is no expectation for a wind.
    \label{fig:vsini_km24}
    }
\end{figure}

\begin{figure} %
     \epsscale{0.8}
     \plotone{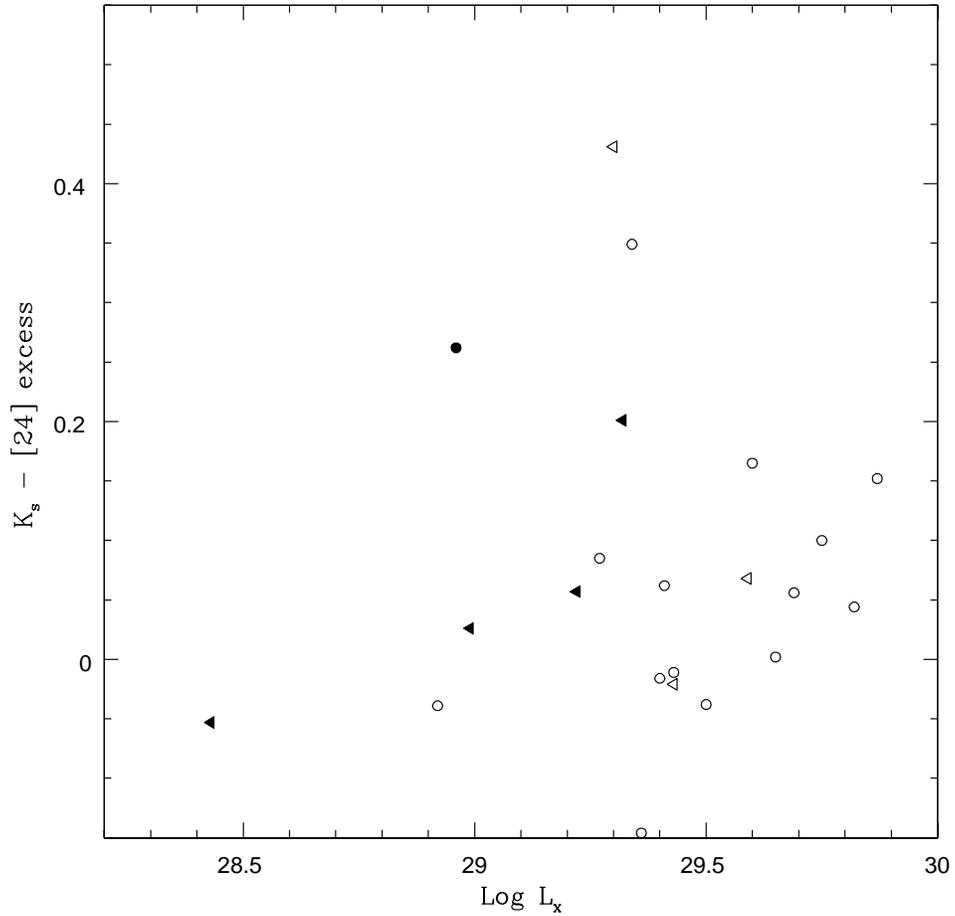}
    \caption{$K{_s}$ - [24] color versus Log (L$_X$) 
for members of Blanco 1 for which we have MIPS data.
    Filled symbols are stars with $B-V$ $<$
  0.35, not expected to be strong x-ray sources (or to have strong
  winds) because they lack outer
  convective envelopes.  Leftward pointing triangles are X-ray upper limits.
    \label{fig:lx_km24}
    }
\end{figure}

\begin{figure} %
     \epsscale{0.8}
     \plotone{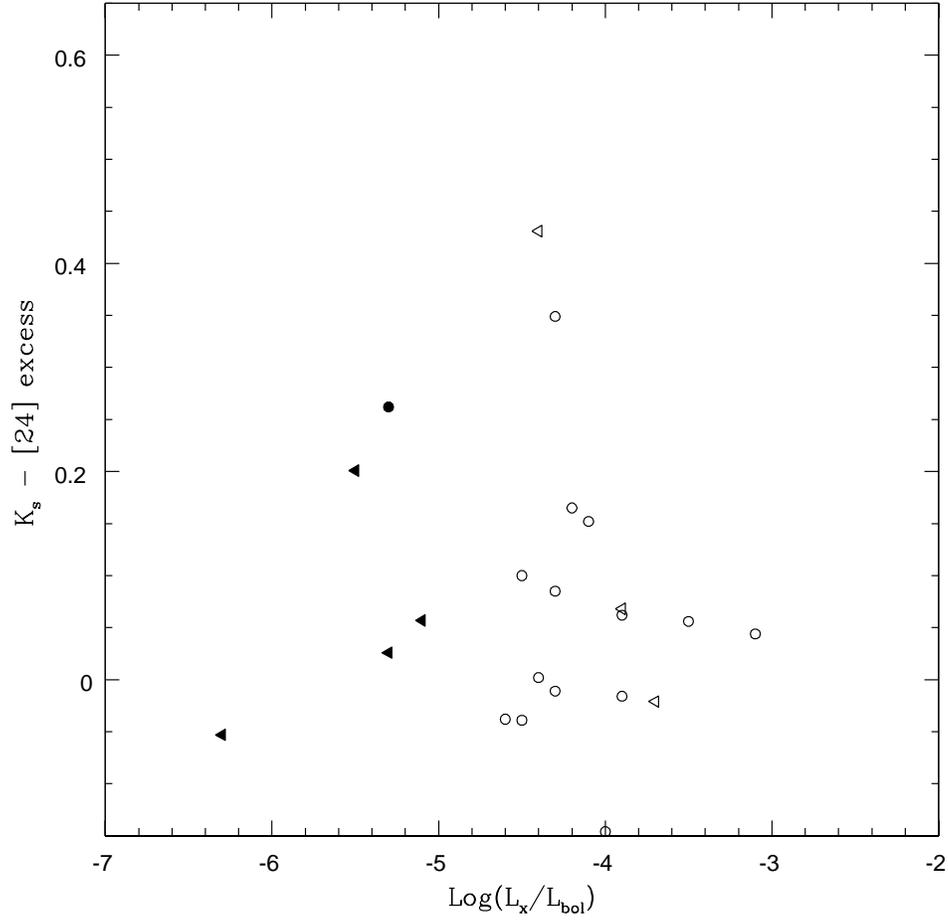}
    \caption{$K{_s}$ - [24] color versus Log (L$_X$/L$_{Bol}$)
for members of Blanco 1 for which we have MIPS data.
    Filled symbols are stars with $B-V$ $<$
  0.35.  Leftward pointing triangles are X-ray upper limits.
    \label{fig:lxlbol_km24}
    }
\end{figure}

There is a theoretical expectation that winds could scour disks of
small particles (Chen et al.\ 2005; Plavchan et  al.\ 2009), and
therefore one might expect a correlation between 24 $\mu$m excess and
a wind proxy.   Figures~\ref{fig:vsini_km24}, \ref{fig:lx_km24}, and
\ref{fig:lxlbol_km24} provide several means to possibly identify such
a correlation.  Stellar rotation (or more specifically differential
rotation) drives the dynamo activity that is thought to power winds
from low mass stars with outer convective envelopes.  The wind mass
loss rate is thought to saturate above some rotational velocity, so
one probably does not expect a linear dependence on rotation - but
at least stars with rotational velocities less than 10-20 \kms\ should
have comparatively weak winds.  Instead, Figure~\ref{fig:vsini_km24}
shows basically no correlation between rotation and IR excess. 
The quiescent X-ray emission of late-type stars arises from non-thermal
heating of the stellar corona, again driven by dynamo activity 
whose ultimate source is differential rotation in the outer convective
envelope.  Above some critical rotational velocity, Log(L$_x$/L$_{Bol}$)
saturates.  It is not known if the wind mass loss rate saturates at
the same critical rotational velocity, but that is a plausible assumption.
If so, then one would expect the largest excesses for stars with
weak x-ray emission and either no excesses or smaller excesses for
stars with saturated x-ray emission.  Figures~\ref{fig:lx_km24} and
\ref{fig:lxlbol_km24} instead show essentially no correlation in the
expected sense.  At face value, this would seem to argue against
the wind scouring model - however, as noted in the next paragraph,
that conclusion is probably unwarranted.  

\begin{figure} %
     \epsscale{0.8}
     \plotone{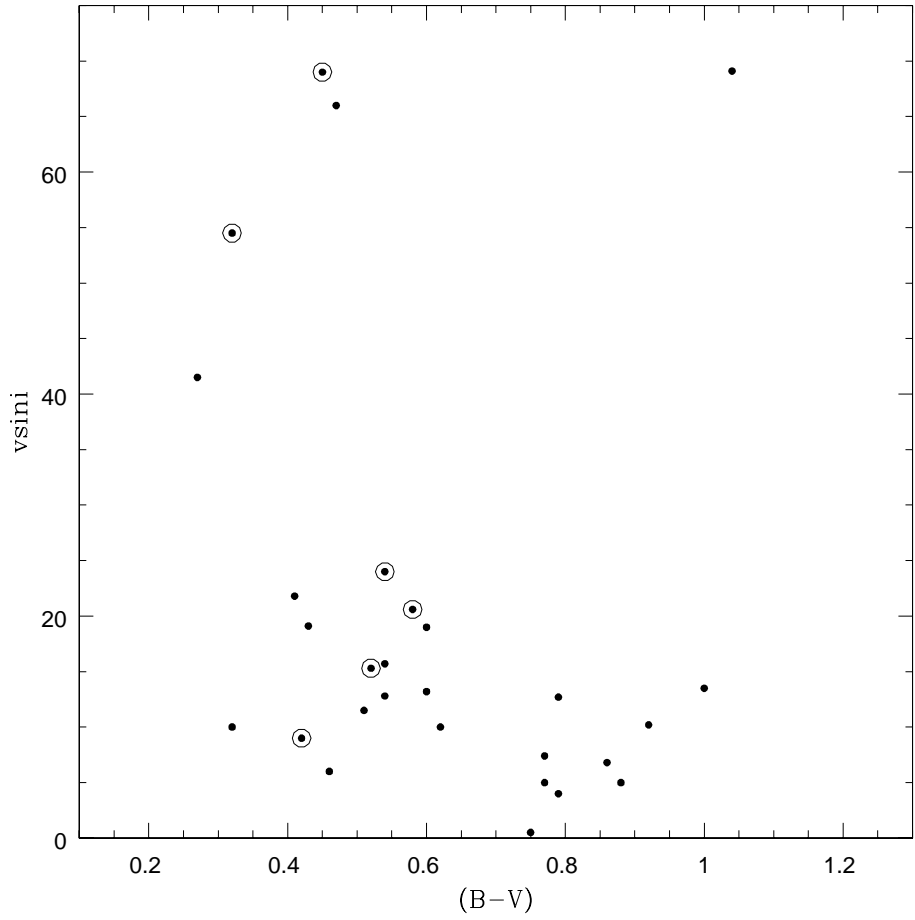}
    \caption{Rotational velocities for Blanco 1 members for which
we have MIPS data.  The encircled dots are stars with apparent 24 $\mu$m
excesses.
    \label{fig:blanco_vsini}
    }
\end{figure}

In Figure~\ref{fig:vsini_km24} we have shown that there is no strong
correlation between rotation and 24 $\mu$m excess for our Blanco 1 FGK
stars.   Is that plot indicative that there truly is no dependence of
mid-IR excess on stellar rotation at this epoch?  We believe the
answer to that question is no because the stars in our sample of
Blanco 1 members are not well-suited to testing this hypothesis.  This
is illustrated in Figure~\ref{fig:blanco_vsini}, a plot of $v \sin i$
as a function of $B-V$ for the Blanco 1 stars for which we have MIPS
data and rotational velocity measures.  The six stars for which we
have both types of data and which have 24 $\mu$m excesses are marked
with open circles. The plot shows that while there is a range in
rotational velocity for the Blanco 1 targets, that range mostly just
reflects the expected dependence of rotation on mass (early F stars
are rapid rotators; late F stars and G dwarfs are slow rotators) --
and at a given mass there is not much spread in rotation.  The stars
with 24 $\mu$m  excesses have rotational velocities typical for their
mass.  Our sample of stars could not provide a strong test of the
dependence of debris disk presence on rotation for stars with outer
convective envelopes because of this.  The only star in our sample
which could provide some degree of a test is ZS38 -- the rapidly
rotating K dwarf with $B-V$ = 1.04 and $v \sin i$ = 69 \kms.  If it had
been found to have a 24 $\mu$m excess, that would have been at odds
with the expectation that a star expected to have a strong dynamo wind
should have scoured its disk.  One star does not provide a good
statistical test.

\begin{figure} %
     \epsscale{0.8}
     \plotone{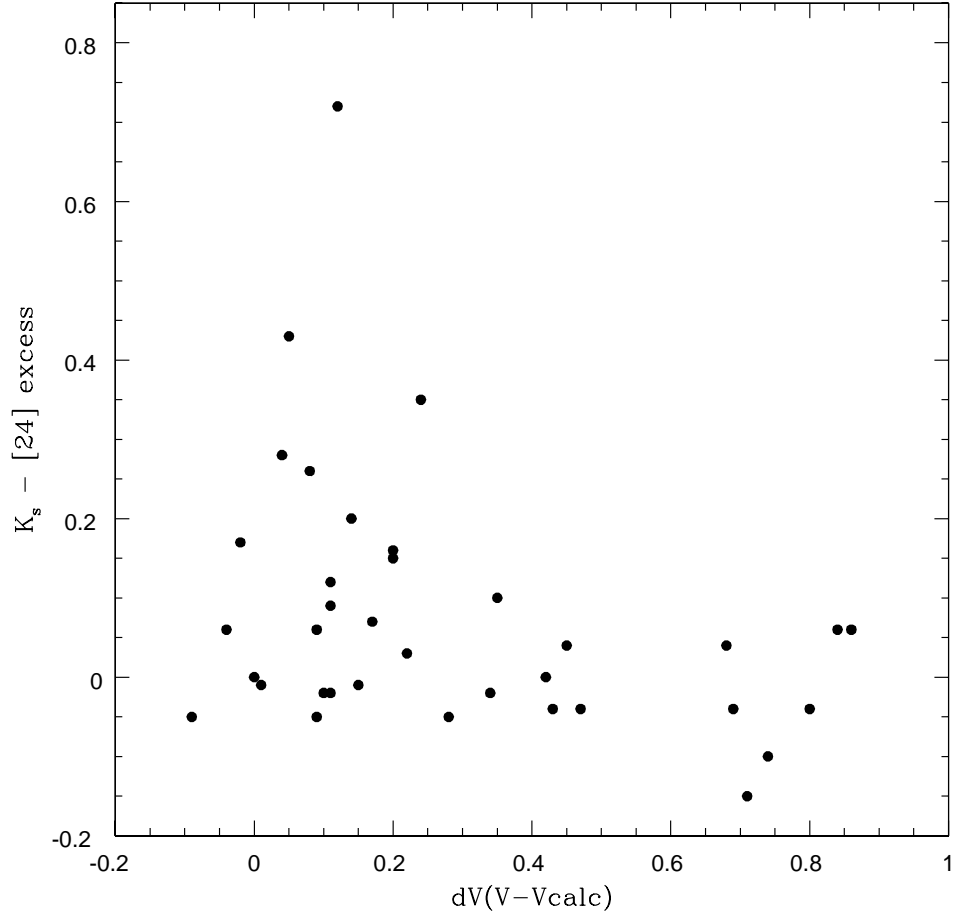}
    \caption{$K_{\rm s} - [24]$ excess versus the height above a single-star
main sequence curve for members of Blanco 1 for which we have MIPS data.
The stars plotted have inferred spectral types from about A0 to mid-K.
    \label{fig:blanco_bin}
    }
\end{figure}

The best empirical linkage between presence of a 24 $\mu$m excess and
another observable which we have been able to identify in our data is
a correlation with a proxy for multiplicity -- specifically, with
displacement above the single-star main sequence in a color-magnitude
diagram.   Figure~\ref{fig:blanco_bin} shows this dependency.  The
displacement above the main sequence has been calculated relative to
the Pleiades single-star main sequence shown in
Figure~\ref{fig:blanco1vmk_cmd}, shifted to Blanco 1 distance.  
Identifying photometric binaries as those stars with displacements
above the single-star main sequence more than 0.3 magnitudes, a
two-sample Kolmogorov-Smirnov test yields just a 6\% chance that the
single and binary stars in Figure~\ref{fig:blanco_bin} have K$_{\rm s}
- [24]$ excess distributions drawn from the same parent distribution. 
We discuss this possible dependence of excess frequency on binarity in
more detail in the next section.

\section{The Correlation between 24$\mu$m Excess and Binarity in Other Open Clusters}

As noted in the preceding section, the strongest linkage between 24
$\mu$m excess and any other stellar property is with a binarity proxy.
Interpretation of that dependency is made more complicated, however,
by the nature of our Blanco 1 object sample.  As shown in
Figures~\ref{fig:blanco1vmk_cmd} and \ref{fig:blanco_km24}, all of our
stars with probable 24 $\mu$m excess have $V-K_{\rm s} <$ 1.5, whereas
most of our stars that are photometric binaries have $V-K_{\rm s} >$
1.5.    Therefore, the apparent link between 24 $\mu$m excess and
displacement above the single star sequence could arise if lower mass
stars in Blanco 1 had much lower probability of having a detectable 24
$\mu$m excess.  A correlation between detected 24 $\mu$m excess and
mass has been noted in some studies (e.g. Siegler et al.\ 2007; Gaspar
et al.\ 2009).  In particular, Figure 6 of Siegler et al.\ indicated a
three times higher 24 $\mu$m excess frequency for B and A stars
compared to FGK dwarfs at the age of the Pleiades (i.e., the age of
Blanco 1). Given the small number of stars in our sample, this could
explain most of the correlation we see in Figure~\ref{fig:blanco_bin}.

A test as to whether the apparent link between 24 $\mu$m excess and
binarity is the result of peculiarities in our sample selection or is
a real effect is to look for a similar correlation in other open
cluster datasets. This is more difficult than it would appear because
(a) one needs a relatively young and rich cluster, so that there are a
significant number of stars with excesses; (b) the cluster needs to
have good membership information and good optical/near-IR photometry;
and (c) the MIPS data must be sensitive enough to detect a large
sample of cluster members (preferably covering the spectral type range
we have sampled in Blanco 1).  We believe that the two clusters that
have the best data to make this test are the Pleiades (age $\sim$ 100
Myr) and NGC 2547 (age $\sim$ 30 Myr, Jeffries and Oliveira 2005).  

For this test, we limit the sample of Pleiades and NGC 2547 stars to
those with 0.0 $< V - K_{\rm s} <$ 3.0 for several reasons: (a) to
approximately match the mass range of the Blanco 1 sample; (b) to
avoid B stars, since young B stars can have IR-excesses from physical
causes other than debris disks (Gorlova et al.\ 2006; 2007); (c)
because the main-sequence slope in V versus $V - K_{\rm s}$ becomes so
steep for $V - K_{\rm s}$ $<$ 0.0 that it becomes very difficult to
determine useful displacements above the main sequence in order to
identify binary stars; and (d) because B stars can be displaced above
the main sequence because they are post-main sequence - also making it
impossible to use their location in a CMD to identify binaries. More
details of the sample of stars in each cluster, their $K_{\rm s} -
[24]$ colors and their displacements above a single star main sequence
can be found in the Appendix.

Figure~\ref{fig:pleiades_bin} and \ref{fig:NGC2547_bin} provide the
plots of $K_{\rm s} - [24]$ excess versus height above the main
sequence for the Pleiades and NGC 2547.   Both plots show the same
dependency as found for Blanco 1, with the photometric binary stars
having a much lower frequency of 24 $\mu$m excesses compared to the
stars that lie near the single-star main sequence. As for Blanco 1, we
use a height above the main sequence of 0.3 magnitude to separate
``single" and binary members of the clusters.  A two-sample KS test
yields just a 2\% chance that the single and binary stars in NGC
2547 have $K_{\rm s} - [24]$ excess distributions drawn from the same
parent distribution; for the Pleiades, the KS test returns a 13\%
probability. Combining all of the single and binary stars for all
three clusters together, the KS test yields a 0.05\% chance that the
single and binary stars are drawn from the same parent distribution. 
Additional discussion of the correlation between binarity and debris
disk detection can be found in Sierchio et al\ (2010).

As shown in Figure~\ref{fig:pleiades_cmd} and \ref{fig:NGC2547_cmd},
neither the Pleiades nor NGC 2547 show an obvious mass dependence for
the  photometric binary frequency.  Taken as a group, these plots for
Blanco 1, the Pleiades, and NGC 2547 provide strong evidence that at
least during the an age range of about 30-100 Myr, for stars of about
1 to 3 $M_{\odot}$ (spectral type A0 to mid-G), there is a strong
relationship between debris disk presence and multiplicity in the
sense that binary stars (or higher multiples) are less likely to have
detected debris disks.  

At face value, our result would seem to be in contradiction to the
Trilling et al.\ (2007) determination that binary stars in the field
were as likely to have detected debris disks as single stars. 
However, there are some notable differences in the two studies that
may vitiate this seeming contradiction. Most importantly, the stars in
the Trilling et al.\ (2007) sample are comparatively old, with all but
three older than 600 Myr and most older than 1 Gyr.  For example, if
winds are a significant scouring agent, they would be much more
effective at the age of our sample than for Trilling's stars.   Second
(but perhaps connected to the first point), most of the excesses for
the Trilling et al.\ (2007) sample were detected at 70 $\mu$m -- only
6 of the 69 stars had excesses detected at 24 $\mu$m, versus 20 of 50
at 70 $\mu$m.    Our sample of stars are too distant to detect at
70 $\mu$m with our observations.   Therefore, we are detecting dust
closer to the parent stars than Trilling (on average). Finally, our
method of identifying binary stars is much different from that used by
Trilling et al.  The two studies are therefore probing somewhat
different star populations and dust  distributions, either of which
may affect the conclusions.

\section{Summary and Conclusions}

Our new MIPS data for Blanco 1 adds one more young cluster to the set
of open clusters with good data from Spitzer.   Because of the
extensive recent optical photometry and high-resolution spectroscopy
obtained by Mermilliod et al.\ (2008), Cargile et al.\ (2009),
Gonzalez \& Levato (2009) and James et al.\ (2010), we have been able
to certify that all of our 25 targeted candidate members are indeed
high probability members of the cluster, and all but one of the
serendipitous members for which we were also able to obtain MIPS
photometry also have enough published ancillary data to confirm their
membership.  Eight of the 25 Blanco 1 cluster members for which we
have MIPS data of good S/N show significant 24 $\mu$m excesses,
corresponding to an excess frequency of 32$\pm$11\%.  This is
comparable to the 24 $\mu$m excess frequency found in the similar-age
Pleiades cluster, and is compatible with the trend of debris disk
frequency versus age reported in several recent studies.  The unusual
galactic orbit and paucity of high mass stars in Blanco 1  appear not
to have had a strong effect on the processes that influence the
formation and evolution of debris disks.

Our most unexpected finding was that we see a much lower IR excess
frequency for photometric binary stars than for stars that lie near
the single-star main sequence for members of Blanco 1, NGC 2547 and
the Pleiades.   This apparent link between binarity and debris disk
presence is counter to what was reported for field A and F stars by
Trilling et al.\ (2007), who found no correlation between binarity and
IR excess.  Because their sample of stars is much older than ours,
there may in fact be no contradiction. The disparity of the two
results, however, does suggest that further studies with different
sample selections would be useful.  Assuming the correlation between
binarity and MIPS 24$\mu$m excess is confirmed, the next step would
be to determine what types of binaries have the most influence -- which
will require samples with binary identification and characterization 
obtained by additional means (radial velocity surveys, AO imaging, etc.).

During the course of our analysis of the Blanco 1 stars, we used MIPS
data for stars in the Hyades to provide an accurate template to define
the photospheric $V-K_{\rm s}$ versus $K_{\rm s}-[24]$ relation for G
and K dwarfs.  As a result of that analysis, we have identified three
Hyades low mass members (vB 19, VA 133 and VA 407) as possible debris
disk sources.

\acknowledgments

This work is based in part on observations made with the Spitzer
Space Telescope, which is operated by the Jet Propulsion Laboratory,
California Institute of Technology under a contract with NASA. Support
for this work was provided by NASA through an award issued by
JPL/Caltech.  Most of the support for this work was provided by the Jet
Propulsion Laboratory, California Institute of Technology, under NASA
contract 1407.   This research has made use of NASA's Star and
Exoplanet Database (NStED) and the Astrophysics
Data System (ADS) Abstract Service, and of the SIMBAD database,
operated at CDS, Strasbourg, France.  This research has made use of
data products from the Two Micron All-Sky Survey (2MASS), which is a
joint project of the University of Massachusetts and the Infrared
Processing and Analysis Center, funded by the National Aeronautics and
Space Administration and the National Science Foundation.  These data
were served by the NASA/IPAC Infrared Science Archive, which is
operated by the Jet Propulsion Laboratory, California Institute of
Technology, under contract with the National Aeronautics and Space
Administration.  The research described in this paper was partially
carried out at the Jet Propulsion Laboratory, California Institute of
Technology, under contract with the National Aeronautics and Space
Administration.  

\appendix

\section{Pleiades and NGC2547 Comparison Sample}

\begin{figure} %
     \epsscale{0.8}
     \plotone{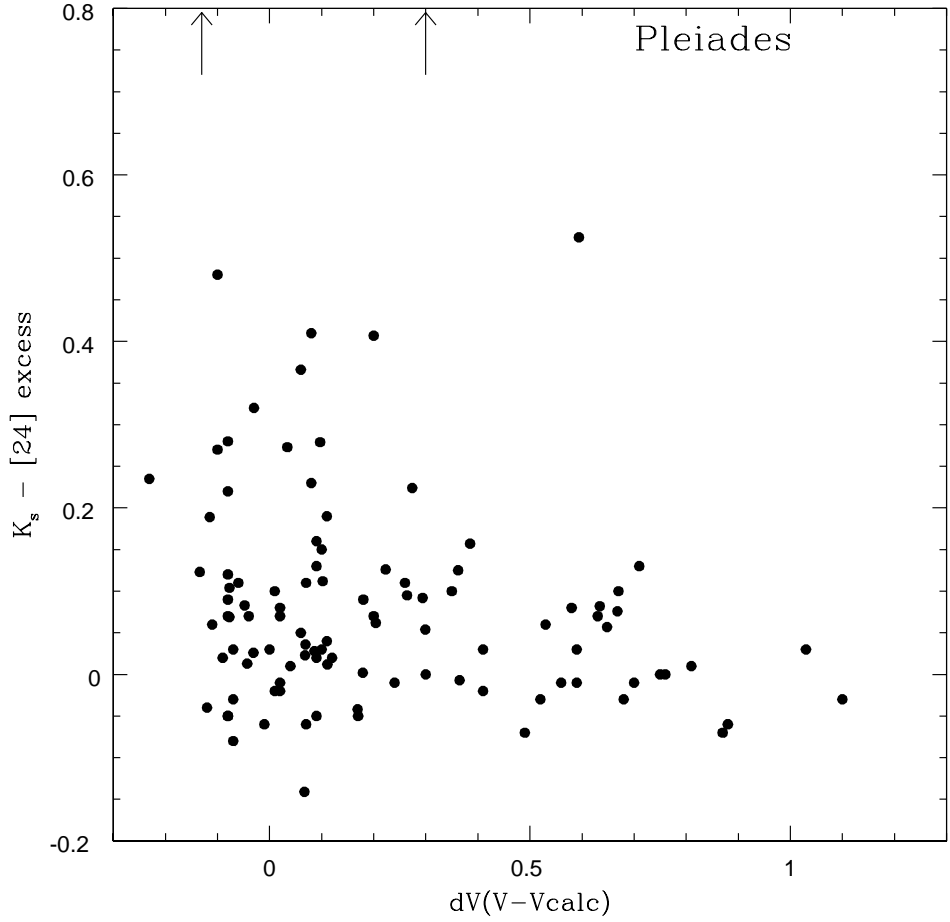}
    \caption{$K_{\rm s} - [24]$ excess versus the height above a single-star
main sequence curve for members of the Pleiades with MIPS data.  Arrows point to
the location of two stars that are off-scale - HII1132 and AKII437.
    \label{fig:pleiades_bin}
    }
\end{figure}

The  Pleiades data used for this comparison  come from Stauffer et
al.\ (2005), Gorlova et al.\ (2006), and Sierchio et al.\ (2010).   We
included all stars from the Stauffer et al. and Sierchio et al. papers
because all of those stars are F, G and K dwarfs that fall well within
our  $V - K_{\rm s}$ color criterion, and all have good S/N
photometry. From the Gorlova et al.\ sample (their Table 2), we
exclude the B stars for the reasons stated in \S 6 and we exclude the
five stars with $V - K_{\rm s}$ $>$ 3.0 (the uncertainties in their
[24] photometry and in the calibration of the photospheric $K_{\rm s}
- [24]$ color make it impossible to know if these stars have excesses
or not). For consistency, for all of the stars we use $V$ and $K_{\rm
s}$ photometry from Stauffer et al.\ (2007) and the 24 $\mu$m
photometry as reported in the above three papers, and then calculate
our own $V - K_{\rm s}$ color and $K_{\rm s} - [24]$ excess and the
displacement relative to the single star main sequence of Stauffer et
al.\ (20007)  using the same procedures as for Blanco 1.  Table 3
provides the data for the stars plotted in Figures
\ref{fig:pleiades_bin} and \ref{fig:pleiades_cmd}.

\begin{figure} %
     \epsscale{0.8}
     \plotone{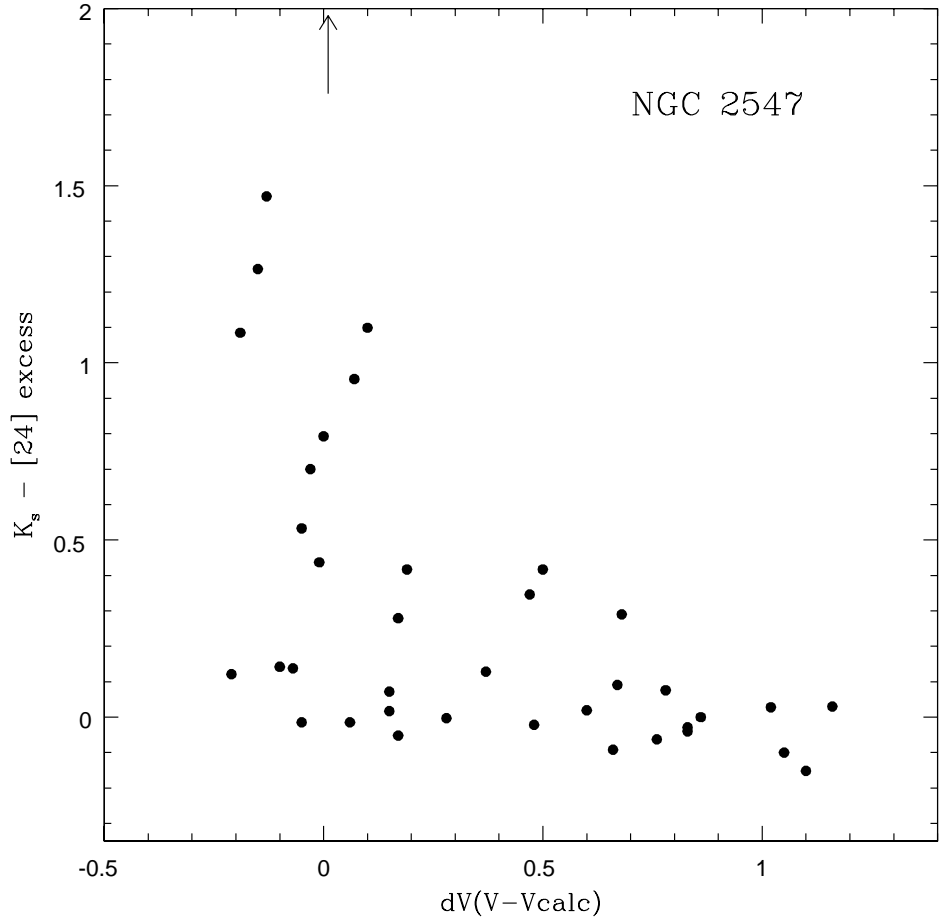}
    \caption{$K_{\rm s} - [24]$ excess versus the height above a single-star
main sequence curve for A0 to mid-K members of NGC 2547 with MIPS data.
The arrow points to the location of one star that is off-scale - ID8
(see Table 5 of Gorlova et al.\ 2007).
    \label{fig:NGC2547_bin}
    }
\end{figure}

\begin{figure} %
     \epsscale{0.8}
     \plotone{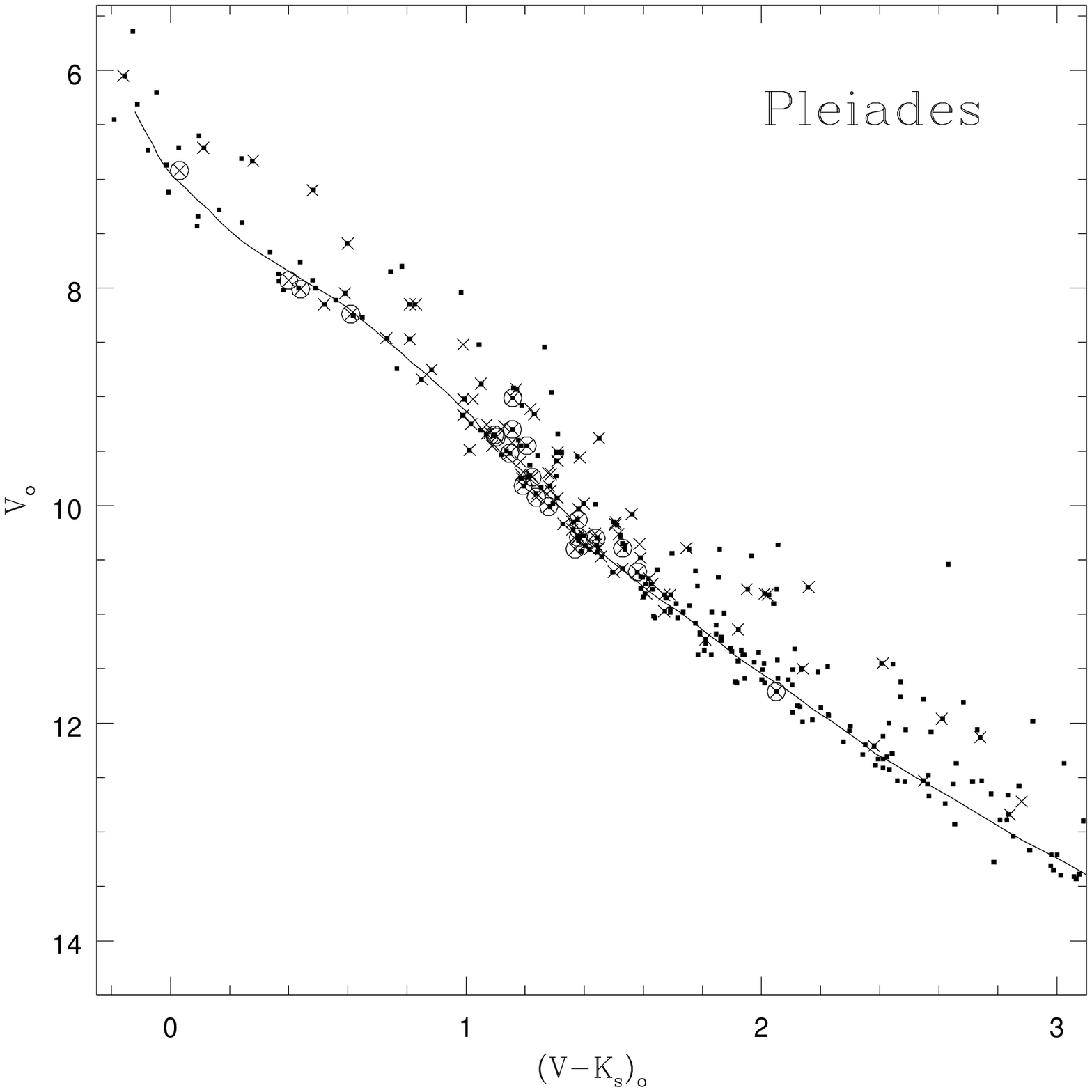}
    \caption{$V_o$ versus $(V-K_{\rm s})_o$ CMD for the Pleiades.  Stars with MIPS 24 $\mu$m
photometry are marked with a cross ($\times$); those with 24 $\mu$m excesses greater
than 15\% of the photospheric flux density are circled.
    \label{fig:pleiades_cmd}
    }
\end{figure}

\begin{figure} %
     \epsscale{0.8}
     \plotone{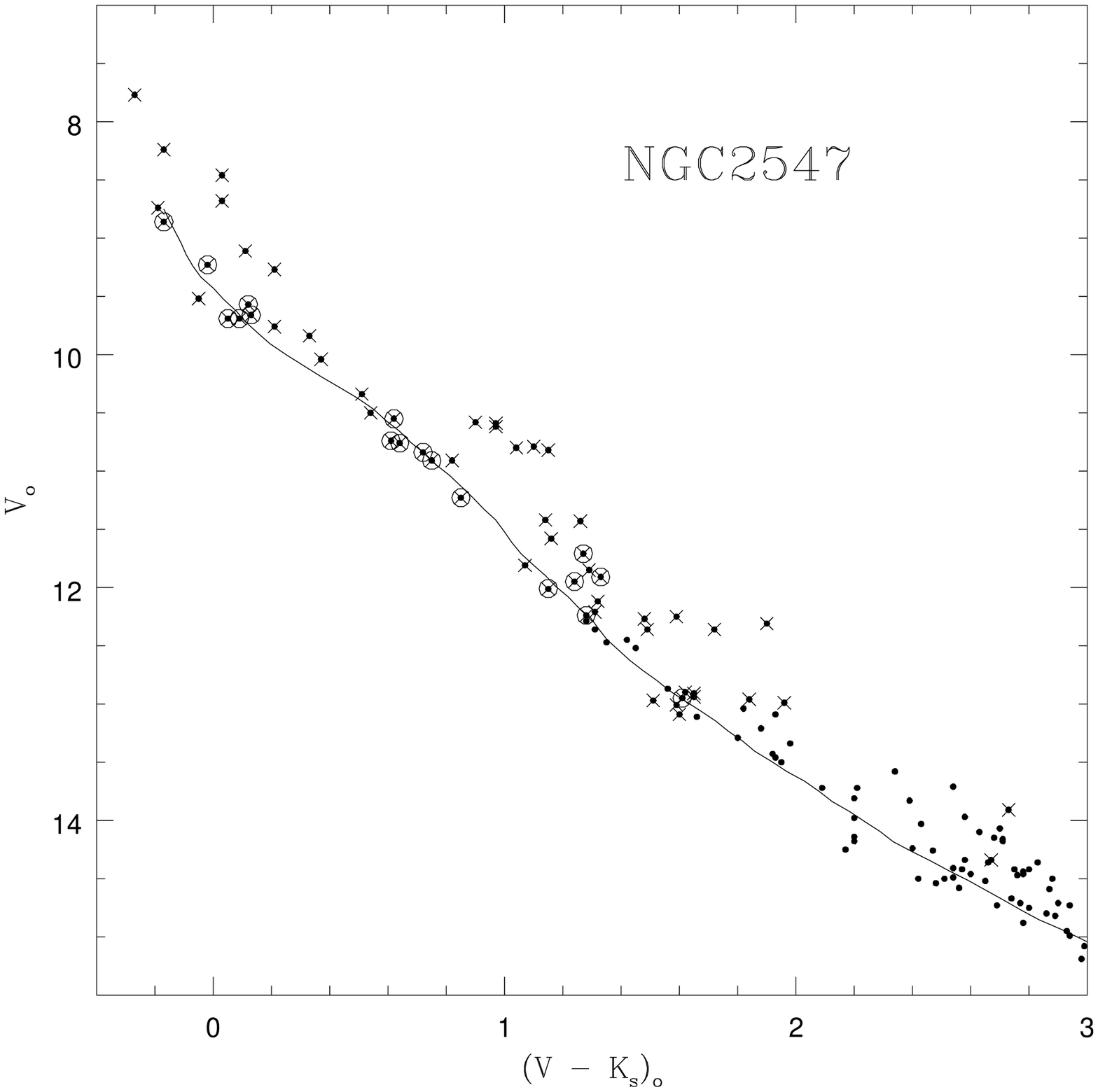}
    \caption{
$V_o$ versus $(V-K_{\rm s})_o$ CMD for NGC 2547.  Stars with MIPS 24 $\mu$m
photometry are marked with a cross ($\times$); those with 24 $\mu$m
excesses greater than 15\% of the photospheric flux density are circled.
    \label{fig:NGC2547_cmd}
    }
\end{figure}

For NGC 2547, we use data from Gorlova et al.\ (2007).  We restrict 
ourselves to the same $V - K_{\rm s}$ range as for the Pleiades, for
the same reasons.  We include candidate cluster members from both
Tables 1 and 2 of Gorlova et al.\ except we eliminate  proper motion
non-members and stars that have 24 $\mu$m data with 1-$\sigma$
uncertainties more than 0.15 mag. We include ID1 and ID8 from the
stars with 8 $\mu$m excesses in Table 5 of Gorlova et al.\ but exclude
the other such stars (ID 2 and 4 have $K$-band excesses, and hence we
cannot use $V - K_{\rm s}$ to identify displacement above the main
sequence; ID 4 has no $K_{\rm s} - [24]$ measurement, and ID 7, which
has $V - K_{\rm s}$ = 5.50).   We include all of the probable 24
$\mu$m excesses in Table 5, except for 08090344-4859210 which has ${(V
- K_{\rm s})_o}$ = 5.37 and is thus nearly 2.5 magnitudes redder than
the reddest Blanco 1 star for which we have MIPS data. We also exclude
the stars listed as just having possible 24 $\mu$m excesses, because
Gorlova et al.\ describe these stars as either having distorted shapes
at 24 $\mu$m or as having 24 flux densities at the level of
fluctuations in the nebulosity.  For those reasons, Gorlova et al.\
did not discuss the sources with ``possible 24 $\mu$m excesses"
further in their paper. Because NGC 2547 is significantly younger than
the Pleiades, we cannot use the Pleiades single-star main sequence
(shifted to NGC 2547 distance) to identify photometric binaries. 
Instead, we use the entire NGC 2547 candidate member list to define
the single star locus for that cluster, and calculate the displacement
relative to that locus.  Table 4 provides the relevant data for the
NGC 2547 stars plotted in Figures \ref{fig:NGC2547_bin} and
\ref{fig:NGC2547_cmd}.

\begin{deluxetable}{lccl}
\tablecaption{Summary of MIPS observations of Blanco 1}
\label{tab:observations}
\tablewidth{0pt}
\tablehead{
\colhead{Target Object} & \colhead{map center (J2000)} &  \colhead{AORKEY} }
\startdata
W23=ZS18  &  0h00m55.35s,-30d03m51.0s & 17690368 \\
W28=ZS26  &  0h01m24.48s,-30d38m58.3s & 17690624 \\
W35=ZS39  & 0h01m57.75s,-30d09m28.7s  & 17690880  \\
W36       & 0h02m04.13s,-30d41m06.8s  & 17691136  \\
W38=ZS48  & 0h02m21.63s,-30d08m21.6s  & 17691392  \\
W51=ZS77  & 0h02m55.23s,-30d08m59.2s  & 17691648 \\
W53=ZS84  & 0h03m10.81s,-30d10m48.9s  & 17691904 \\
W56=ZS91  & 0h03m20.61s,-29d49m22.7s  & 17692160 \\
W57=ZS96 & 0h03m21.85s,-30d01m10.5s  & 17696512 \\
W58=ZS90  & 0h03m24.39s,-29d48m49.3s  & 17692416 \\
W60=ZS101 & 0h03m28.98s,-30d19m27.2s  & 17692672 \\
W61=ZS104 & 0h03m31.89s,-29d43m04.8s  & 17692928 \\
W63=ZS99  & 0h03m33.59s,-30d28m42.0s  & 17693184 \\
W68=ZS107 & 0h03m50.17s,-30d03m55.7s  & 17693440 \\
W70=ZS110 & 0h04m00.02s,-30d19m15.8s  & 17693696 \\
W79=ZS129 & 0h04m31.66s,-30d14m41.6s  & 17695744 \\
W86=ZS136 & 0h04m50.82s,-29d37m58.9s  & 17693952 \\
W88=ZS139 & 0h04m53.33s,-30d15m23.8s  & 17694208 \\
W91=ZS138 & 0h04m58.84s,-30d09m41.6s  & 17694464 \\
W96=ZS157 & 0h05m26.76s,-30d17m24.5s  & 17694720 \\
W99=ZS 160 & 0h05m30.94s,-29d53m08.2s  & 17696256 \\
W104=ZS166 & 0h05m42.96s,-29d57m38.2s  & 17694976 \\
W113=ZS182 & 0h06m16.35s,-30d05m57.0s  & 17696000 \\
W124=ZS199 & 0h07m07.81s,-30d33m41.6s  & 17695232 \\
W126       & 0h08m01.90s,-30d01m54.1s  & 17695488 \\
\enddata
\end{deluxetable}

\begin{deluxetable}{lccccccccccc}
\tablecaption{Blanco 1 MIPS and Ancillary Data}
\tabletypesize{\tiny}
\rotate
\tablewidth{0pt}
\tablehead{
\colhead{Star Name\tablenotemark{a}} &  \colhead{RA (2000)} &
\colhead{Dec (2000) } & \colhead{$V$} &
\colhead{$B-V$} & \colhead{$K_{\rm s}$} & \colhead{[24]} & \colhead{$K_{\rm s} - [24]$} &
\colhead{$v \sin i$} & \colhead{Log(L$_x$)} & \colhead{Log(L$_x$/L$_{Bol}$)} & \colhead{Comments\tablenotemark{c}}\\
& \colhead{(degrees)} & \colhead{(degrees)} & & & & & &
\colhead{(km/s)} & \colhead{(erg/s/cm$^2$/Hz)} & &  }
\startdata
W23 = ZS18* & 0.230577 & -30.064169 &  8.39 &  0.04 &  8.355 &  7.64(0.01) &  0.71 & \nodata& \nodata & \nodata & G09;NS \\
W28 = ZS26* & 0.351998 & -30.649542 & 10.50 &  0.40 &  9.484 &  9.20(0.02) &  0.28 &   9.0 &  \nodata & \nodata & M08 \\
W35 = ZS39* & 0.490606 & -30.157967 &  9.96 &  0.32 &  9.158 &  8.87(0.02) &  0.27 &  54.5 &  28.96P & -5.3 & G09; P78   \\
W36 = M901 & 0.517190 & -30.685226 &  8.98 &  0.10 &  8.715 &  8.71(0.01) & -0.01 & \nodata & \nodata& \nodata & G09; P78  \\
W38 = ZS48* & 0.590125 & -30.139343 & 10.72 &  0.48 &  9.601 &  9.43(0.03) &  0.16 &  69.0 &  29.87C & -4.10 & M08 \\
W51 = ZS77 & 0.730105 & -30.149778 &  8.43 &  0.02 &  8.427 &  8.48(0.01) & -0.05 &  \nodata& $<$28.43P & -6.3 & G09; P78; NS   \\
W53 = ZS84* & 0.795057 & -30.180246 & 11.32 &  0.56 &  9.967 &  9.79(0.04) &  0.18 &  24.0 &  29.60C & -4.2 & M08 \\
W56 = ZS91 & 0.835889 & -29.822985 & 11.30 &  0.57 &  9.961 &  9.97(0.05) &  0.00 &  12.8 &  29.43C & -4.3 & M08 \\
W57 = ZS96 & 0.841037 & -30.019594 & 10.38 &  0.42 &  9.271 &  9.30(0.02) & -0.03 &  21.8 &  29.50C & -4.6 & M08; SB \\
W58 = ZS90 & 0.851605 & -29.813707 & 10.62 &  0.50 &  9.462 &  9.41(0.02) &  0.01 &  66.0 &  29.65C & -4.4 & M08 \\
W60 = ZS101 & 0.870738 & -30.324226 & 10.61 &  0.44 &  9.553 &  9.42(0.03) &  0.13 &  19.1 &  \nodata& \nodata & M08  \\
W61 = ZS104 & 0.882868 & -29.718004 & 10.05 &  0.36 &  9.127 &  9.02(0.02) &  0.11 &   \nodata&  29.75C & -4.5 & G09; C09 \\
W63 = ZS99 & 0.889975 & -30.478338 & 10.62 &  0.45 &  9.498 &  9.46(0.02) &  0.04 &   6.0 &   \nodata& \nodata  & M08 \\
W68 = ZS107 & 0.959037 & -30.065468 & 11.04 &  0.54 &  9.844 &  9.77(0.03) &  0.07 &  11.5 &  28.95C & -4.9 & M08; SB2 \\
W70 = ZS110 & 1.000072 & -30.321060 & 11.12 &  0.56 &  9.764 &  9.75(0.04) & -0.03 &  15.7 &  \nodata & \nodata  & M08; SB \\
W79 = ZS129 & 1.131912 & -30.244907 & 11.68 &  0.61 & 10.232 & 10.15(0.05) &  0.08 &   7.1 & $<$29.59M & $<$-3.9 & M08  \\
W86 = ZS136 & 1.211755 & -29.633030 &  8.29 &  0.01 &  8.273 &  8.31(0.01) & -0.05 &  \nodata &  \nodata & \nodata & G09; P78; NS  \\
W88 = ZS139* & 1.222222 & -30.256613 &  8.32 &  0.05 &  8.271 &  8.07(0.01) &  0.20 &  \nodata & $<$29.32M & $<$-5.5 & G09; P78; NS  \\
W91 = ZS138* & 1.245159 & -30.161558 & 11.47 &  0.58 & 10.082 &  9.72(0.03) &  0.36 &  18.3 &  29.34C & -4.3 & M08 \\
W96 = ZS157 & 1.361498 & -30.290138 &  9.73 &  0.27 &  9.072 &  9.01(0.02) &  0.06 &  41.5 & $<$29.22M & $<$-5.1 & G09; P78  \\
W99 = ZS160* & 1.378900 & -29.885614 & 11.26 &  0.55 &  9.993 &  9.55(0.03) &  0.44 &  15.3 & $<$29.30M & $<$-4.4 & M08 \\
W104 = ZS166 & 1.429001 & -29.960606 &  9.92 &  0.32 &  9.122 &  9.08(0.02) &  0.03 & $<$10   & $<$28.99M & $<$-5.3  & G09; P78  \\
W113 = ZS182 & 1.568121 & -30.099182 & 11.72 &  0.63 & 10.249 & 10.14(0.05) &  0.10 &   8.0 &  29.27C & -4.3 & M08 \\
W124 = ZS199 & 1.782528 & -30.561554 &  8.67 &  0.09 &  8.538 &  8.53(0.01) &  0.00 & \nodata &  \nodata &  & G09; P78; NS  \\
W126 = ZS226 & 2.007917 & -30.031666 &  9.78 &  0.31 &  9.038 &  9.04(0.02) & -0.01 &  \nodata &   \nodata &  & G09; P78; NS  \\
ZS38\tablenotemark{b}   & 0.476847 & -30.128284 & 13.86 &  1.03 & 10.932 & 10.80(0.09) &  0.13 &  69.1 &  29.82C & -3.1 & J99; C09; SB2 \\
ZS54\tablenotemark{b}   & 0.617458 & -30.078751 & 13.03 &  1.00 & 10.510 & 10.40(0.08) &  0.11 &  13.5 &  29.69C & -3.5 & M08; VB \\
ZS62\tablenotemark{b}   & 0.647750 & -30.117222 & 12.56 &  0.77 & 10.561 & 10.55(0.09) &  0.01 &   5.0 &  29.40C & -3.9 & M08; SB \\
ZS83\tablenotemark{b}   & 0.779556 & -30.254759 & 12.51 &  0.85 & 10.380 & 10.49(0.09) & -0.11 &   6.8 &  29.36C & -4.0 & C09 \\
ZS95\tablenotemark{b}   & 0.818703 & -29.979839 & 12.42 &  0.92 & 10.266 & 10.34(0.08) &  0.00 &  10.2 &  28.92C & -4.5 & M08 \\
ZS100\tablenotemark{b}   & 0.863876 & -30.446903 & 12.55 &  0.79 & 10.507 & 10.44(0.09) &  0.07 &  12.7 &  \nodata &  \nodata & M08; SB  \\
ZS102\tablenotemark{b}   & 0.890083 & -30.262501 & 12.50 &  0.77 & 10.729 & 10.57(0.09) &  0.19 &   7.4 &  \nodata &  \nodata & M08 \\
ZS108\tablenotemark{b}   & 0.977224 & -30.283806 & 13.41 &  1.06 & 10.747 & 10.72(0.09) &  0.03 & \nodata &  \nodata &  \nodata & J10; SB2  \\
ZS161\tablenotemark{b}   & 1.361920 & -29.855742 & 12.67 &  0.79 & 10.754 & 10.75(0.09) &  0.00 &   4.0 & $<$29.43M & $<$-3.7 & M08  \\
ZS165\tablenotemark{b}   & 1.398048 & -29.951780 & 12.49 &  0.88 & 10.260 & 10.16(0.08) &  0.10 &   5.0 &  29.41C & -3.9 & M08; SB \\
W89\tablenotemark{b}     & 1.244333 & -29.563276 & 11.55 &  0.59 & 10.094 & 10.14(0.08) & -0.04 &  19.0 & \nodata & \nodata & M08  \\
M348\tablenotemark{b} & 0.854329 & -30.435089 & 11.97 &  0.75 & 10.080 & 10.16(0.08) & -0.08 &   0.5 & \nodata & \nodata & M08; SB  \\
\enddata
\tablenotetext{a}{The stars with an asterisk (*) after their name have 24 $\mu$m excesses,
probably due to their having debris disks, according to our analysis.}
\tablenotetext{b}{Serendipitously imaged member.}
\tablenotetext{c}{C09: Cargile et al.\ 2009; G09: Gonzalez \& Levato 2009; J99: Jeffries \& James 1999;
J10: James et al.\ 2010; M08: Mermilliod et al.\ 2008; P78: Perry et al.\ 1978}
\end{deluxetable}

\begin{deluxetable}{lccccccccccc}
\tablecaption{Pleiades Data}
\tabletypesize{\tiny}
\rotate
\tablewidth{0pt}
\tablehead{
\colhead{Star Name} &  \colhead{$V - K_{\rm s}$} &
\colhead{$K_{\rm s} - [24]$ exc. } & \colhead{$\delta V$} &
\colhead{Star Name} &  \colhead{$V - K_{\rm s}$} &
\colhead{$K_{\rm s} - [24]$ exc. } & \colhead{$\delta V$} &
\colhead{Star Name} & \colhead{$V - K_{\rm s}$}  & \colhead{$K_{\rm s} - [24]$ exc.} &
\colhead{$\delta V$} }
\startdata
 AK1A317 & 1.18 &  0.02 &  0.07 & HII0717 & 0.48 & -0.06 &  0.88 & HII1876 & 0.28 &  0.01 &  0.88 \\ 
 AK1A36  & 1.52 &  0.05 &  0.30 & HII0727 & 1.31 &  0.03 &  0.41 & HII1912 & 1.17 & -0.01 &  0.41 \\ 
 AK1A56  & 1.59 &  0.12 &  0.36 & HII0739 & 1.45 &  0.03 &  1.03 & HII1924 & 1.36 &  0.07 &  1.03 \\ 
 AK1A76  & 1.10 &  0.28 &  0.10 & HII0885 & 2.61 &  0.10 &  0.67 & HII2027 & 2.01 &  0.00 &  0.67 \\ 
 AK1B146 & 1.31 & -0.07 &  0.49 & HII0923 & 1.38 &  0.09 &  0.18 & HII2034 & 2.55 & -0.02 &  0.18 \\ 
 AK1B365 & 1.28 &  0.06 &  0.20 & HII0996 & 1.38 &  0.22 & -0.08 & HII2147 & 2.16 & -0.03 & -0.08 \\ 
 AK1B560 & 1.15 &  0.37 &  0.06 & HII1015 & 1.42 & -0.03 & -0.07 & HII2172 & 1.37 &  0.12 & -0.07 \\ 
 AK1B590 & 1.02 & -0.04 &  0.17 & HII1095 & 2.05 &  0.28 & -0.08 & HII2195 & 0.44 &  0.27 & -0.08 \\ 
 AK1B7   & 1.29 &  0.13 &  0.22 & HII1100 & 2.74 &  0.13 &  0.71 & HII2278 & 2.02 &  0.00 &  0.71 \\ 
 AK1B8   & 1.38 &  0.07 & -0.08 & HII1101 & 1.38 &  0.41 &  0.08 & HII2311 & 1.81 & -0.08 &  0.08 \\ 
 AKII34  & 1.22 &  0.06 &  0.65 & HII1117 & 1.56 &  0.08 &  0.58 & HII2341 & 1.61 &  0.07 &  0.58 \\ 
 AKII359 & 1.46 &  0.01 & -0.04 & HII1122 & 0.99 &  0.11 & -0.06 & HII2345 & 0.99 & -0.05 & -0.06 \\ 
 AKII383 & 1.24 &  0.19 & -0.12 & HII1132 & 1.16 &  3.08 &  0.30 & HII2506 & 1.36 & -0.06 &  0.30 \\ 
 AKII437 & 1.19 &  1.47 & -0.13 & HII1139 & 1.02 &  0.10 & -0.08 & HII2644 & 1.67 & -0.05 & -0.08 \\ 
 AKIII28 & 1.16 &  0.00 &  0.18 & HII1182 & 1.44 & -0.01 &  0.02 & HII2786 & 1.33 & -0.04 &  0.02 \\ 
 HCG131  & 2.84 & -0.05 &  0.17 & HII1200 & 1.28 &  0.13 &  0.09 & HII2881 & 2.41 & -0.07 &  0.09 \\ 
 HCG132  & 2.88 &  0.10 &  0.35 & HII1207 & 1.40 &  0.02 & -0.09 & HII3031 & 0.88 &  0.03 & -0.09 \\ 
 HII0025 & 1.10 &  0.15 &  0.10 & HII1266 & 0.83 & -0.01 &  0.56 & HII3097 & 1.69 &  0.03 &  0.56 \\ 
 HII0102 & 1.75 &  0.08 &  0.63 & HII1284 & 0.61 &  0.32 & -0.03 & HII3179 & 1.31 & -0.06 & -0.03 \\ 
 HII0120 & 1.63 &  0.02 &  0.09 & HII1298 & 2.38 &  0.05 &  0.06 & Pels7   & 1.44 &  0.01 &  0.06 \\ 
 HII0152 & 1.50 &  0.12 & -0.08 & HII1309 & 1.07 & -0.02 &  0.02 & Pels20  & 1.37 &  0.23 &  0.02 \\ 
 HII0173 & 1.95 & -0.03 &  0.68 & HII1338 & 0.99 &  0.03 &  0.59 & Pels23  & 1.40 &  0.09 &  0.59 \\ 
 HII0174 & 2.14 &  0.00 &  0.30 & HII1362 & 0.52 &  0.06 & -0.11 & Pels25  & 1.09 &  0.03 & -0.11 \\ 
 HII0250 & 1.53 &  0.10 &  0.01 & HII1380 & 0.03 &  0.19 &  0.11 & Pels40  & 1.29 &  0.04 &  0.11 \\ 
 HII0293 & 1.62 &  0.02 &  0.12 & HII1384 & 0.60 & -0.01 &  0.59 & Pels58  & 0.40 &  1.01 &  0.59 \\ 
 HII0314 & 1.59 & -0.01 &  0.24 & HII1431 & 0.11 &  0.06 &  0.53 & Pels86  & 1.07 &  0.11 &  0.53 \\ 
 HII0344 & 0.59 &  0.04 &  0.11 & HII1514 & 1.45 &  0.08 &  0.02 & Pels121 & 1.50 & -0.01 &  0.02 \\ 
 HII0405 & 1.22 &  0.01 &  0.04 & HII1613 & 1.19 &  0.03 & -0.07 & Pels124 & 1.18 &  0.08 & -0.07 \\ 
 HII0489 & 1.44 &  0.23 &  0.08 & HII1726 & 1.23 &  0.07 &  0.63 & Pels128 & 1.51 &  0.16 &  0.63 \\ 
 HII0514 & 1.58 &  0.16 &  0.09 & HII1762 & 0.81 & -0.03 &  0.52 & Pels135 & 1.13 &  0.09 &  0.52 \\ 
 HII0530 & 0.85 &  0.09 & -0.08 & HII1766 & 1.16 &  0.52 &  0.59 & Pels146 & 1.53 &  0.41 &  0.59 \\ 
 HII0531 & 0.73 &  0.07 &  0.02 & HII1776 & 1.67 &  0.11 &  0.07 & Pels150 & 1.21 &  0.22 &  0.07 \\ 
 HII0571 & 1.92 &  0.11 &  0.26 & HII1794 & 1.40 &  0.03 &  0.00 & Pels173 & 1.01 &  0.16 &  0.00 \\ 
 HII0605 & 1.05 & -0.02 &  0.41 & HII1797 & 1.28 &  0.48 & -0.10 & Pels174 & 1.38 &  0.08 & -0.10 \\ 
 HII0697 & 0.81 &  0.07 &  0.20 & HII1856 & 1.24 & -0.05 & -0.08 & Tr60    & 1.08 & -0.14 & -0.08 \\ 
\enddata
\end{deluxetable}

\begin{deluxetable}{lccccccc}
\tablecaption{NGC 2547 Data}
\tabletypesize{\scriptsize}
\rotate
\tablewidth{0pt}
\tablehead{
\colhead{Star Name} &  \colhead{$V - K_{\rm s}$} &
\colhead{$K_{\rm s} - [24]$ exc. } & \colhead{$\delta V$} &
\colhead{Star Name} & \colhead{$V - K_{\rm s}$}  & \colhead{$K_{\rm s} - [24]$ exc.} &
\colhead{$\delta V$} }
\startdata
 08090250-4858172 & 1.61 &  3.64 &  0.01 & 08111134-4904442 & 0.71 &  0.53 & -0.05 \\
 08095601-4919299 & 0.03 & -0.10 &  1.05 & 08095066-4912493 & 0.37 &  0.02 &  0.15 \\
 08100607-4914180 & 0.05 &  1.47 & -0.13 & 08094610-4914270 & 0.54 & -0.01 & -0.05 \\
 08100841-4900434 & 0.13 &  1.10 &  0.10 & 08102667-4906532 & 0.51 & -0.01 &  0.06 \\
 08084571-4923473 & 0.62 &  0.95 &  0.07 & 08091401-4904029 & 0.85 &  0.14 & -0.10 \\
 08093671-4911383 & 0.03 & -0.04 &  0.83 & 08093053-4920443 & 0.82 &  0.07 &  0.15 \\
 08112585-4912288 & 0.09 &  0.70 & -0.03 & 08101836-4906461 & 1.27 &  0.42 &  0.50 \\
 08092668-4914371 & 0.59 &  1.26 & -0.15 & 08102774-4912095 & 1.15 & -0.15 &  1.10 \\
 08100087-4908324 & 1.04 &  0.00 &  0.86 & 08101165-4922274 & 1.26 & -0.06 &  0.76 \\
 08102082-4903366 & 0.11 &  0.02 &  0.60 & 08104984-4911258 & 1.16 &  0.13 &  0.37 \\
 08103144-4906301 & 0.21 & -0.09 &  0.66 & 08085576-4923085 & 1.33 &  0.35 &  0.47 \\
 08110323-4900374 & 0.57 &  1.09 & -0.19 & 08085649-4923128 & 1.59 &  0.29 &  0.68 \\
 08093053-4921563 & 0.33 &  0.00 &  0.28 & 08100938-4900565 & 1.90 &  0.03 &  1.16 \\
 08101673-4915173 & 0.72 &  0.79 &  0.00 & 08103983-4904377 & 1.14 & -0.02 &  0.48 \\
 08104233-4857253 & 0.12 &  0.28 &  0.17 & 08101352-4920438 & 1.10 &  0.03 &  1.02 \\
 08104662-4917312 & 0.90 &  0.09 &  0.67 & 08094507-4856307 & 1.28 &  0.44 & -0.01 \\
 08110860-4900161 & 0.21 & -0.05 &  0.17 & 08104546-4901068 & 1.24 &  0.42 &  0.19 \\
 08085310-4913492 & -.05 &  0.12 & -0.21 & 08103432-4900496 & 1.07 &  0.14 & -0.07 \\
 08093561-4927015 & 0.97 & -0.03 &  0.83 & 08101546-4905487 & 1.72 &  0.08 &  0.78 \\
\enddata
\end{deluxetable}

\vfill
\eject







\clearpage

\clearpage
\newpage
\markright{Figure 3}

\begin{thebibliography}{}


\bibitem[Adams et al.(2004)]{adams04}
Adams, F. et al. 2004, \apj, 611, 360

\bibitem[Balog et al.(2009)]{balog09}
Balog, Z. et al. 2009, \apj, 698, 1989

\bibitem[Bate et al.(2003)]{bate03}
Bate, M. et al. 2003, \mnras, 339, 577

\bibitem[Beichman et al.(2006)]{beichman06}
Beichman, C. et al. 2006, \apj, 652, 1674

\bibitem[Bertoldi (1989)]{bertoldi89}
Bertoldi, F. 1989, \apj, 346, 735

\bibitem[Bessell (2000)]{bessell00}
Bessell, M. 2000, \pasp, 112, 961

\bibitem[Bryden et al.(2006)]{bryden06}
Bryden, G. et al.  2006, \apj, 636, 1098

\bibitem[Cargile et al.(2009)]{cargile09}
Cargile, P., James, D.J., \& Platais, I.  2009, \aj, 137, 3230

\bibitem[Carpenter et al.(2008)]{carpenter08}
Carpenter, J. et al. 2008, \apjs, 179, 423

\bibitem[Carpenter et al.(2009)]{carpenter09}
Carpenter, J. et al. 2009, \apjs, 181, 197

\bibitem[Chen et al.(2005)]{chen05}
Chen, C., Jura, M., Gordon, K., \& Blaylock, M.  2005, \apj, 623, 493

\bibitem[Cieza et al.(2008)]{cieza08}
Cieza, L., Cochran, W., \& Augereau, J.-C.  2008, \apj, 679, 720

\bibitem[Currie et al.(2008)]{currie08}
Currie, T., Plavchan, P., \& Kenyon, S.  2008, \apj, 688, 597

\bibitem[Dias et al.(2002)]{dias02}
Dias, W., Alessi, B., Moitinho, A., \& Lepine, J.  2002, \aap, 389, 871

\bibitem[Edvardsson et al.(1995)]{edvardsson95}
Edvardsson, B. et al.  1995, \aap, 293, 75

\bibitem[Engelbracht et al.(2007)]{engelbracht07}
Engelbracht, C. et al. 2007, \pasp, 119, 994

\bibitem[de Epstein \& Epstein(1985)]{epstein85}
de Epstein, A. \& Epstein, I.  1985, \aj, 90, 1211

\bibitem[Gaspar et al.(2009)]{gaspar09}
Gaspar, A., et al.  2009, \apj, 697, 1578

\bibitem[Gautier et al.(2007)]{gautier07}
Gautier, N. et al.\ 2007, \apj, 667, 527

\bibitem[Gonzalez \& Levato(2009)]{gonzalez09}
Gonzalez, J. \& Levato, H. 2009, \aap, 507, 541

\bibitem[Gorlova et al.(2004)]{gorlova04}
Gorlova, N. et al.\  2004, \apjs, 154, 448

\bibitem[Gorlova et al.(2006)]{gorlova06}
Gorlova, N. et al.\  2006, \apj, 649, 1028

\bibitem[Gorlova et al.(2007)]{gorlova07}
Gorlova, N. et al.\  2007, \apj, 670, 516

\bibitem[Hauschildt et al.(1999)]{hauschildt99}
Hauschildt, P. et al.\ 1999, \apj, 512, 377

\bibitem[James et al.(2010)]{james10}
James, D. J. et al.\ 2010, in preparation

\bibitem[Jeffries \& James(1999)]{jeffries09}
Jeffries, R.D., \& James, D.  1999, \apj, 511, 218.

\bibitem[Jeffries \& Oliveira(2005)]{jeffries05}
Jeffries, R.D., \& Oliveira, J.  2005, \mnras, 358, 13.

\bibitem[Kenyon \& Hartmann(1995)]{kenyon95}
Kenyon, S. \& Hartmann, L.  1995, \apjs, 101, 117

\bibitem[Kenyon \& Bromley(2004)]{kenyon04}
Kenyon, S. \& Bromley, B.  2004, \apj, 602, L133

\bibitem[Lada et al.(1991)]{lada91}
Lada, E. et al.  1991, \apj, 371, 171

\bibitem[Makovoz \& Marleau(2005)]{makovoz05}Makovoz, D., \&
Marleau, F. 2005, \pasp, 117, 1113

\bibitem[Mermilliod et al.(2008)]{mermilliod08}
Mermilliod, J.-C., Platais, I., James, D.J., Grenon, M., \&
Cargile, P.  2008, \aap, 485, 95

\bibitem[FEPS)]{FEPS}
Meyer, M., et al.\ 2006, \pasp, 118, 1690

\bibitem[Micela et al.(1999)]{micela99}
Micela, G. et al.  1999, \aap, 344, 83

\bibitem[Moraux et al.(2007)]{moraux07}
Moraux, E., Bouvier, J., Stauffer, J., Barrado y Navascues, D., \&
Cuillandre, J.-C.  2007, \aap, 471, 499

\bibitem[Motoyama et al.(2007)]{motoyama07}
Motoyama, K., et al. 2007, \aap, 467, 657

\bibitem[Oort \& Spitzer(1955)]{oort55}
Oort, J. \& Spitzer, L.  1955, \apj, 121, 60

\bibitem[Panagi et al.(1994)]{panagi94}
Panagi, P., et al. 1994, \aap, 292, 439

\bibitem[Panagi \& O'Dell(1997)]{panagi97}Panagi, P., \&
O'Dell, M. 1997, \aaps, 191, 213

\bibitem[Papovich et al.(2004)]{papovich04}
Papovich, C. et al.  2004, \apjs, 154, 70

\bibitem[Perry et al.(1978)]{perry78}
Perry, C.L., Walter, D.K., \& Crawford, D.L. 1978, \pasp, 90, 81

\bibitem[Perryman et al.(1997)]{perryman97}
Perryman, M. et al.  1997, \aap, 323, L49

\bibitem[Pillitteri et al.(2003)]{pillitteri03}
Pillitteri, I. et al. 2003, \aap, 399, 919

\bibitem[Pillitteri et al.(2004)]{pillitteri04}
Pillitteri, I. et al. 2004, \aap, 421, 175

\bibitem[Plavchan et al.(2005)]{plavchan05}
Plavchan, P. et al. 2005, \apj, 631, 1161

\bibitem[Plavchan et al.(2009)]{plavchan09}
Plavchan, P. et al. 2009, \apj, 698, 1068

\bibitem[Porras et al.(2003)]{porras03}
Porras, A. et al.  2003, \aj, 126, 1916

\bibitem[Rebull et al.(2008)]{rebull08}
Rebull, R., et al., 2008, ApJ, 681, 1484

\bibitem[mips]{mips}Rieke, G., et al., 2004, ApJS, 154, 25

\bibitem[Rieke et al.(2005)]{rieke05}
Rieke, G. et al.\  2005, \apj, 620, 1010

\bibitem[Schmidt-Kaler. (2005)]{schmidtkaler82}
Schmidt-Kaler, T.H.\  1982, Physical Parameters of the Stars, in
   Landolt Bornstein New Series, Vol. 2b.

\bibitem[Siegler et al.(2007)]{siegler07}
Siegler, N., Muzerolle, J., Young, E., Rieke, G., Mamajek, E., Trilling, D.,
  Gorlova, N., Su, K.  2007, \apj, 654, 580

\bibitem[Sierchio et al.(2010)]{sierchio10}
Sierchio, J. M. et al.\ 2010, \apj, 712, 1421

\bibitem[Skrutskie et al.(2006)]{skrutskie06}
Skrutskie, M. et al.\ 2006, \aj, 131, 1163

\bibitem[Stauffer et al.\ (2005)]{stauffer05}
Stauffer, J.~R., et al.\  2005, \aj, 130, 1834

\bibitem[Stauffer et al.\ (2007)]{stauffer07}
Stauffer, J.~R., et al.\  2007, \apjs, 172, 663

\bibitem[Su et al.(2006)]{su06}
Su, K. et al.\  2006, \apj, 653, 675

\bibitem[Su et al.(2009)]{su09}
Su, K. et al. 2009, in preparation

\bibitem[Trilling et al.(2007)]{trilling07}
Trilling, D. et al.\  2007, \apj, 658, 1289

\bibitem[Werner et al.(2004)]{werner04}
Werner, M., et al., 2004, ApJS, 154, 1

\bibitem[Westerlund et al(1988)]{westerlund88}
Westerlund, B., Garnier, R., Lundgren, K., Pettersson, B.,
\& Breysacher, J. 1988, \aaps, 76, 101

\end{thebibliography}
\end{document}